\documentclass[rmp,aps,amssymb,twocolumn,nofootinbib,floatfix]{revtex4}

\usepackage[ruled]{algorithm2e}

\SetAlFnt{\small}
\SetAlCapFnt{\small}
\SetAlCapNameFnt{\small}
\SetAlCapHSkip{0pt}
\IncMargin{-\parindent}

\usepackage{graphicx}
\usepackage{amsmath,amssymb}
\usepackage[ruled]{algorithm2e}
\usepackage{multirow}
\usepackage{array}
\usepackage{algpseudocode}
\usepackage{color}
\usepackage{url}
\newtheorem{definition}{Definition}

\begin{document}


\title{Community Discovery in Dynamic Networks: a Survey}

\author{Giulio Rossetti}
\email{giulio.rossetti@di.unipi.it}
\affiliation{Knowledge Discovery and Data Mining Lab, ISTI-CNR, Pisa, Italy}
 
\author{Remy Cazabet}
\email{remy.cazabet@gmail.com}
\affiliation{Sorbonne Universites, UPMC Univ Paris 06, CNRS, LIP6 UMR 7606}
\affiliation{Univ Lyon, UCBL, CNRS, LIRIS UMR 5205, F-69621, Lyon, France}

\begin{abstract}
Several research studies have shown that Complex Networks modeling real-world phenomena are characterized by striking properties: (i) they are organized according to community structure and (ii) their structure evolves with time.
Many researchers have worked on methods that can efficiently unveil substructures in complex networks, giving birth to the field of community discovery.
A novel and fascinating problem started capturing researcher interest recently: the identification of evolving communities. 
Dynamic networks can be used to model the evolution of a system: nodes and edges are mutable and their presence, or absence, deeply impacts the community structure that composes them.

This survey aims to present the distinctive features and challenges of dynamic community discovery and propose a classification of published approaches.
As a ``user manual", this work organizes state of the art methodologies into a taxonomy, based on their rationale, and their specific instantiation.
Given a definition of network dynamics, desired community characteristics and analytical needs, this survey will support researchers to identify the set of approaches that best fit their needs. 
The proposed classification could also help researchers to choose in which direction to orient their future research.
\end{abstract}

\keywords{Dynamic Networks, Temporal Networks, Community Discovery}
\maketitle

\tableofcontents


\section{Introduction}
Complex Networks \cite{newman2003structure} are popular tools -- both theoretical and analytical -- commonly used to describe and analyze interaction phenomena that occur in the real world. 
Social ties formation, economic transaction, face to face communication, the unfolding of human mobility are a few examples of observable events that are often investigated using instruments borrowed from Graph Theory.

One main issue to address while studying such kind of real-world events lies in the identification of meaningful substructures hidden within the overall complex system.
How to partition a complex network into \emph{communities} is a widely studied problem \cite{fortunato2010community,Coscia2011}: several works, from a broad set of disciplines, proposed different community definitions and approaches able to retrieve such structures automatically.
The magnitude of papers that cover this fascinating topic highlights its ubiquity: 
communities can capture groups of strongly connected online documents, individuals sharing a social environment as well as products being frequently purchased together. 

Most of the scientific literature dedicated to the community discovery problem start from a very stringent assumption: real-world phenomena can be modeled with static networks, i.e., mathematical objects \emph{``frozen"} in time.
Unfortunately, such simplified scenario seldom fits the evolving nature of the world we live in. 
Indeed, one of the key features that characterize interaction phenomena is that they naturally unfold through time: social interactions evolve, phone calls have a limited duration, human mobility changes as time goes by.
The temporal dimension conveys highly valuable information that the analyst needs to exploit to better understand the reality he/she observes. 
The need to include such knowledge in the tools offered by graph theory has been one of the linchpins in an entirely new field of investigation that emerged in the last decade: dynamic network analysis.
Within this new field, well-known problems defined in static settings were reformulated, and time-aware approaches able to solve them were proposed.

Amongst them, Dynamic Community Discovery (henceforth also referred as \textbf{DCD}) represents one of the most challenging problems. 
As time goes by, nodes and edges can join and leave a dynamic network, producing relevant perturbations on its topology: communities are the basic bricks of that complex structure and, as such, are profoundly affected by local changes.
While classical community discovery algorithms aim to identify hidden substructures, dynamic approaches are devoted to the tracking of such local topologies and of their mutations, a goal that can be fulfilled pursuing different strategies and modeling choices.

In this survey, we have thus chosen to design a taxonomy of DCD algorithms not by considering the technical solution they use to find them, but rather the strategy they use to identify time-aware meaningful network substructures.
Our aim is to support researchers and analysts in choosing the approach, or at least the class of approaches, that better fits their needs and data.
For each of the proposed classes of DCD algorithms we discuss both  advantages and drawbacks. Moreover, we provide a brief description of those methods that implement their rationale.

To provide a comprehensive view of this field of research, we also discuss issues related to the evaluation of DCD algorithms, which requires modifications compared with the static case; moreover, we summarize other works on DCD, such as dynamic community visualization methods, and analytical applications to real-world problems.

The survey is organized as follows. 
In Section \ref{sec:dyn_network}, we discuss the general concepts behind dynamic network analysis and the mathematical models used to describe, in graph theoretical terms, temporal evolving phenomena.
In Section \ref{sec:dynamic_cd}, we formalize the Community Discovery problem on Dynamic networks and discuss community life-cycle and instability.
In Section \ref{sec:classification}, we introduce our DCD taxonomy describing the peculiarity, pros, and cons of each class of approaches. A detailed description of the methods is then made available in the \textbf{Appendix}.
In Section \ref{sec:evaluation} we introduce and discuss two classes of techniques often used to compare and characterize communities identified by DCD algorithms, namely \emph{internal} and \emph{external} quality evaluation.
In Section \ref{subsec:realdata} we review some real-world scenarios that have been studied through the lenses of dynamic community mining and address the problem of visualizing analytical results.
Finally Section \ref{sec:conclusion} concludes the survey, providing insights into DCD related open issues.
\label{sec:intro}


\section{Dynamic Networks}
\label{sec:dyn_network}
Since its beginning, complex network analysis has been approached by researchers through the definition of very specific mining problems. 
Among them 
community discovery \cite{fortunato2010community,Coscia2011}, link-based object ranking \cite{getoor2005link,duhan2009page}, frequent pattern mining \cite{milo2002network,alon2007network} are examples of analytical tasks originally defined on networks \emph{``frozen in time"}. 

The absence of the time dimension comes from historical reasons: the graph theory ancestry of the field, and the scarcity of existing dynamic data at the time the field of complex networks emerged.

With the explosion of human-generated data, often collected by socio-technical platforms, more and more datasets having rich temporal information that can be studied as dynamic networks becomes available.
Many works tried to transpose well-known graph mining problems from static networks to temporal networks: Temporal motifs mining \cite{kovanen2011temporal}, Diffusion \cite{lee2012exploiting}, Link prediction \cite{tabourier2016predicting}, etc.

To better address the central topic of this survey, we first introduce in Section \ref{subsec:rep} the most used formalisms to represent \emph{dynamic networks} and, subsequently in Section \ref{subsec:memory}, we discuss one of the leading issues arising when dealing with temporally annotated networks: \emph{network memory}.

\subsection{Network Models}
\label{subsec:rep}
Time plays a crucial role in shaping network topologies. 
One of the most important issue to address in order to reason on time evolving networks is thus related to the mathematical formalism used to describe them.
\begin{figure}[!th]
\centering
\includegraphics[width=1\columnwidth]{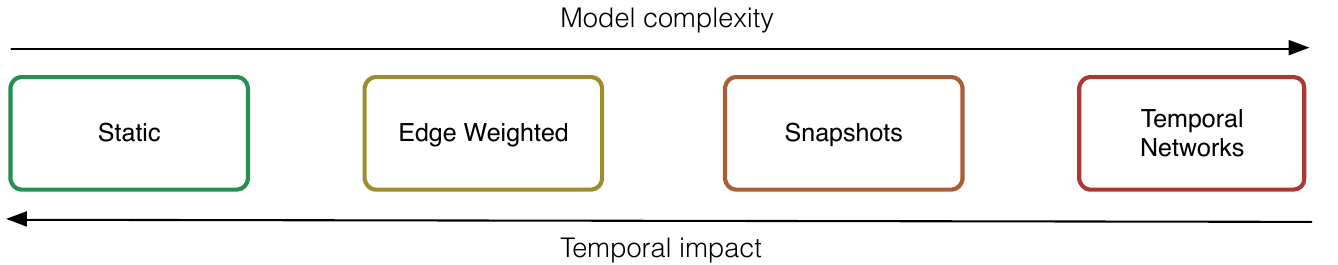}
\caption[Network representations]{Network representations. Moving from \emph{static} graph to \emph{interaction} ones the model complexity increases while the grain of the temporal information available decreases \cite{rossetti2015social}.}
\label{img:time_complexity}
\end{figure}
In Figure \ref{img:time_complexity} are shown four different conceptual solutions that gradually introduce the temporal dimension in the network modelling process. 
At one hand we find the complete \emph{atemporal} scenario (i.e., a single network snapshot capturing a static glimpse of a dynamic phenomenon): network dynamics can be gradually introduced by adopting labels that weights nodes and edges, thus capturing an \emph{aggregate} network. 
Following this approach, weights models the number of occurrence of a given node/edge in a pre-determined observation window: with this little increment in the model complexity, several time-dependent analysis neglected before become possible (i.e., tie strength estimation).
Aggregation strategies, however, suffer a severe limitation, they do not capture dynamics. 
For this reason several works model dynamic phenomena with temporally ordered series of network \emph{snapshots}.
This simple modeling choice allows to efficiently keep track of the perturbations occurring on the network topology. 
However, while the expressivity of the model is increased, the analytical complexity increases as well.
When dealing with network snapshots to perform time-aware mining tasks, two issues need to be addressed: (i) how to keep track of multiple stages of the network life and (ii) how to harmonize the analytical results obtained in a snapshot with the outcome of subsequent ones.

Dynamic network partition, as well as aggregation, suffers from a pressing issue: to be performed, a temporal granularity -- a threshold to transform the original dynamic system -- needs to be fixed. 
Since the identification of such threshold is not trivial --  it is, indeed, context-dependent and often profoundly impacts analytical results -- recent works propose to model dynamic phenomena without any aggregation, keeping all temporal details. 
Such studies usually decompose a dynamic network in its elementary bricks: temporally ordered, timestamped, \emph{interactions} or relations. We will describe in more details such approach, namely \textit{temporal networks} modeling, in the next section.

Temporal networks, avoiding aggregation, allow for a complete and fine-grained description of network dynamics: as a drawback, this solution increases the complexity of the network model and requires the definition of novel analytical methodologies. 

\bigbreak
Different problems, as well as available data, impose different modeling choices: static network, as well as weighted ones, are often used to identify stable patterns and to describe the actual status of a network while snapshots and interactions are proxies for the study of increasingly dynamic scenarios. 
Obviously, starting from fine-grained temporal data, it is possible to generate all the other models by subsequent aggregations.
Since we are interested in community detection approaches that deals with temporally annotated graphs, in the following we will highlight the major strengths and drawbacks of the most used models in this context: \textbf{Temporal Networks} and \textbf{Network Snapshots}. 
In this survey, we will not discuss static and weighted networks since, by definition, they do not allow to make direct use of temporal information, thus reducing dynamic problems to static ones: for an in-depth overview of community discovery approaches defined for such scenarios refer to \cite{fortunato2010community,fortunato2016community,xie2013overlapping,Coscia2011}.

\subsubsection{Temporal Networks}
Several formalisms have been proposed to represent evolving networks without loss of information: Temporal Networks \cite{Holme2013}, Time-Varying Graphs \cite{casteigts2012time}, Interaction Networks \cite{rossetti2015interaction}, Link Streams and Stream graphs \cite{VLM15,latapy2017stream}, to name the most popular. Hereafter, we use the term \textit{Temporal Networks}, with a general definition that encompasses all those formalisms.

A Temporal Network models a dynamic structure in which both nodes and edges may appear and disappear as time goes by, more formally:

\begin{definition}[Temporal Network]
A Temporal Network is a graph $G=(V,E,T)$ where: $V$ is a set of triplets of the form $(v,t_s,t_e)$, with $v$ a vertex of the graph and $t_s, t_e \in T$ are respectively the birth and death timestamps of the corresponding vertex (with $t_s \leq t_e$); $E$ is a set of quadruplets $(u,v,t_s,t_e)$, with $u, v \in V$ are vertices of the graph and $t_s, t_e \in T$ are respectively the birth and death timestamps of the corresponding edge (with $t_s \leq t_e$).
\end{definition}
Depending on the interaction semantics, we can deal with undirected Temporal Networks (TN) or with Directed Temporal Networks (DTN).

The high-level definition proposed encompasses nodes and edges with or without duration (if $t_s = t_e$).
Often, a strong distinction is made between those two categories. Networks without edges durations are often referred as \textit{contact sequences}, those with durations as \textit{interval graphs} \cite{Holme2013}. 
We argue that, for community detection, this distinction is not relevant to characterize the nature of temporal networks. 
Indeed,  a \textit{contact sequence} can always be transformed into an \textit{interval graph} by extending edges of an infinitesimal amount. 
Conversely, an \textit{interval graph} can be  discretized into a \textit{contact sequence} at the desired resolution. 
However, such transformations will not fundamentally change the nature of the network. 

Differently, a distinction must be made between two types of temporal networks, \textit{interaction networks} and \textit{relation networks}: the former model interactions either with duration (phone calls, face to face communication, etc.) or not (emails, short messages, etc.) that can repeat as time goes by. 
Relation networks, conversely, model more stable relations (friendship, co-worker, belonging to the same group, etc.). 
In relation networks, the state of the graph at any given time is well-defined and can be studied through classical static analysis tools.
Conversely, in interaction networks, the analyst needs to focus his attention on a temporal windows or on aggregations of some sort to obtain a \textit{stable} dynamic graph that does not change dramatically between close time-periods. 
This distinction is important since many methods -- in particular the ones working on edge streams -- update the state of communities after each atomic network perturbation. 
Indeed, classical approaches do not produce interpretable results when edges are instantaneous or have very low persistence (e.g., phone calls). 

Due to the lack of literature, and in particular of a formal definition able to distinguish between these two types of temporal networks, in this article we have no choice but to ignore this distinction. 
We stress that most algorithms present in the current survey implicitly assume a relation network, save three methods \cite{matias2018semiparametric,VLM15,HMNS16}, specifically designed to work with edges without duration.



\subsubsection{Network Snapshots}

Often, network history is partitioned into a series of snapshots, each one of them corresponding either to the state of the network at a time $t$ (relation network) or to the aggregation of observed interactions during a period (interaction network). 
\begin{definition}[Snapshot Network]
A Snapshot Graph $\mathcal{G}_{\tau}$ is defined by an \textbf{ordered} set $\langle G_1,G_2 \dots G_t \rangle$ where each snapshot $G_i=(V_i,E_i)$ is univocally identified by the sets of nodes $V_i$ and edges $E_i$.
\end{definition}
Depending on the network semantic, we can deal with undirected Snapshot Graphs (SN) or with Directed Snapshot Graphs (DSN). 
Dynamic networks represented using snapshots are similar in nature to multiplex (or multislice) networks, which are \textbf{unordered} sets of static networks. To learn more on community detection on multislice networks, readers can refer to \cite{doi:10.1093/comnet/cnu016}. When the adaptation of methods to the dynamic case have been explored, methods will be included in the present article, see for instance \cite{Mucha2010}.

Network snapshots can be effectively used, for instance, to model a phenomenon that generates network perturbations (almost) at regular intervals. 
In this scenario, context-dependent temporal windows are used to partition the network history into consecutive snapshots: time-bounded observations describing a precise, static, discretization of the network life. 

The snapshot-based analysis is frequently adopted for networks having a natural temporal discretization, due to the balance they propose between model complexity and expressivity. 
SN allow to apply static networks tools on evolving networks. 
Algorithms, as well as network measures and definitions, can be independently applied to each temporal partition without the need of novel analytical tools that explicitly manage temporal information.
SN can also be seen as a convenient modeling choice to avoid the instability problem of \textit{temporal} networks, using aggregation over periods of times and sliding windows \cite{morini2017revealing}. 

However, the strength of this model represents also its major drawback.
Even if network snapshots reduce the analytical complexity and allow for task parallelization, they intrinsically imply the need for a reconciliation phase in which the independently computed, partial results are combined. 
The need for \emph{two-step} (mining and reconciliation) procedures can often be the cause of the quality degradation of analytical results. 
Another issue related to the SN model is the identification of the optimal window size to use when generating the snapshots, a choice that can profoundly affect the outcome of the subsequent analysis. 

\subsubsection{Discussion: Temporal Networks vs. Snapshots}
An important issue to address is the relation between \emph{Snapshot} and \emph{Temporal Networks} since, as we will see in Section \ref{sec:classification}, a generic DCD algorithm is usually tailored to exploit only one of such representations. 
The differences among TN and SN are tightly connected to the constraints such models impose: the choice of employing TN instead of SN (or vice versa) can have significant implication on aspects related to the storage of data, and the processing part of the designed analysis. 
In particular:

\begin{itemize}
    \item[a)]If the original data already describe punctual states of the network evolution (for instance, a weekly/monthly/yearly complete crawl of a system), the snapshot model is obviously the natural one to adopt. 
    On the contrary, if a more precise temporal information is available (for instance an e-mail, a friend list or phone call dataset with timestamps of finite precision), both solutions can be considered. 
    \item[b)] Since, unlike TNs, SNs are composed of aggregated data, the typical complexity of processes on a SN depends on the chosen level of aggregation, and on the sizes of the produced snapshots (number of nodes/edges). 
    For instance, methods for DCD on SN typically need to run a community detection process on each snapshot. 
    The cost of running a Community Discovery algorithm at $t+1$ typically does not depend on the number of changes between SN at $t$ and $t+1$. 
    Processes running on TNs, on the contrary, typically depend mainly on the number of network modifications, not on the granularity or the total number of nodes. 
    \end{itemize}
\subsection{Network Memory}
\label{subsec:memory}
A modeling choice that profoundly affects the analysis performed on dynamic networks regards the system \emph{memory}. 
When the analyzed data describe an interaction network (for instance, e-mails or phone calls), not composed of long-lasting relations, most methods require to transform it into a relation network, i.e., to assign to each edge a duration appropriate to consider it as relation network. 
Such transformation can yield snapshots (if synchronized) or temporal networks (if each edge duration computation is done independently).

Two main scenarios are often considered: (i) \emph{perfect} memory network (or accumulative growth scenario) and (ii) \emph{limited} memory network.

In the former scenario nodes and edges can only join the network: the dynamic network has a perfect memory meaning that old edges/nodes cannot disappear (e.g., a citation graph). 
Conversely, in the latter, nodes/edges can vanish as time goes by (i.e., in a social network context, edge disappearance can be used to model the decay of interactions' social effect). 

We call \textit{Time To Live} (TTL) the artificially defined duration of an edge. Several strategies exist to fix this TTL:
\begin{itemize}
    \item \emph{fixed size static time window}: the TTL is equal for all network entities, and there is a finite set of possible periods whose start and end dates are defined in advance. This strategy produces snapshot networks.  
    \item \emph{fixed size sliding time window}: the TTL is equal for all network entities, but it starts independently for each one of them at the moment of their first appearance (i.e., an email graph in which each email is considered to have the same social persistence starting from the time it is made);
    \item \emph{dynamic size time window}: the TTL is equal for all the network entities at a given time but depends on the current network status (i.e., a system for which it is interesting to capture bursts of interactions having different frequency);
    \item \emph{(global/local) decay function}: the TTL of each node/edge is defined independently, usually as a function of the global/local network activity (i.e., fitting the inter-arrival time distribution of nodes or edges for the whole network as well as for the individual node pair).
\end{itemize}
The assumptions made on the persistence of network entities (e.g., the strategy identified to fix their TTL) play a crucial role in the results provided by time-aware mining algorithms.

\section{Dynamic Community Discovery}
\label{sec:dynamic_cd}
As discussed in the previous section, the temporal dimension profoundly impacts the analytical tasks that are commonly performed on complex networks.
Among them, a problem that has always received high attention from the scientific community is community discovery (henceforth, {\bf CD}). 
Networks built on real-world data are complex objects often composed of several substructures hidden in an apparent chaos: community discovery aims to decompose them in meaningful sub-topologies that better describe local phenomena. 
Even though the problem itself is intuitively clear, its formulation is particularly ill-posed: given a specific network, different partitions can exist, each of them capturing different but valuable information. 
In the last decades, we have witnessed the proliferation of an extensive literature on such subject, each work proposing its approach optimizing the partition for a different quality function (i.e., modularity, density, conductance, cut\dots), thus making the comparison between different partitions challenging. 
The widespread of different community definitions lead to a renewed awareness: the perfect Community Discovery Algorithm does not exist. Conversely, there exist a set of algorithms that better perform on a specific declination of the general problem. 

In this section, we formally define the community discovery problem on dynamic graphs (Section \ref{subs:problem_def}) and describe how its solutions can be used to study community life-cycle (Section \ref{subset:lifecycle}). 
Moreover, we discuss two relevant topics related to this complex problem: community instability and temporal smoothing (Section \ref{sec:instability}).

\subsection{Problem Definition}
\label{subs:problem_def}
Community discovery is a typical problem in complex network analysis. 
One of the main reason behind its complexity is undoubtedly related to its ill-posedness. 
Indeed, several community definitions were proposed so far. 
Classic works intuitively describe communities as sets of nodes closer among them than with the rest of the network, while others, looking at the same problem from another angle, only define such topologies as dense network subgraphs. 
To maintain a more general perspective, in this survey we will adopt a meta-definition -- borrowed from \cite{Coscia2011} -- to create an underlying concept able to generalize to all the variants found in the literature (verbatim from \cite{Coscia2011}): 

\begin{definition}[Community] A community in a complex network is a set of entities that share some closely correlated sets of actions with the other entities of the community. We consider a direct connection as a particular and very important, kind of action.
\end{definition}

Given the high-level nature of the adopted problem definition, one could prefer the term of ``node clustering" over community detection. 
Due to the lack of consensus in the literature, we choose to keep the term \emph{communities} that is specific to the field of networks mining and analysis.

Once fixed the definition of community, we can formally define the specific task of identifying and tracking mesoscale structures in a dynamic network scenario. 
Even in this case, we adopt a generic definition, that does not make any assumption about the nature of communities, and that correspond to any dynamic cluster of nodes.
\begin{definition}[Dynamic Community Discovery]
 Given a dynamic network $DG$, a Dynamic Community $DC$ is defined as a set of distinct (node, periods) pairs: $DC= \{ (v_1,P_1),(v_2,P_2), \dots ,(v_n,P_n) \} $, with $P_n =((t_{s0},t_{e0}), (t_{s1},t_{e1})\dots (t_{sN}, t_{eN}))$, with $t_{s*} \leq t_{e*} $.  Dynamic Community Discovery aims to identify the set $\mathcal{C}$ of all the dynamic communities in $DG$. The partitions described by $\mathcal{C}$ can be neat as well as overlapping.
\end{definition}

Within this general definition fall approaches having different goals. 
In particular, two main analytical aims can be identified: (i) enable for an efficient identification of optimal partitions for each instant of the evolution and, (ii) build evolutive chains describing the life-cycle of communities.


\subsubsection{Community Life-Cycle}
\label{subset:lifecycle}
The persistence along time of communities subjected to progressive changes is an important problem to tackle. 
As illustrated by the famous paradox of the ship of Theseus, deciding if an element composed of several entities at a given instant is the same or not as another one composed of some -- or even none -- of such entities at a later point in time is necessarily arbitrary, and cannot be answered unambiguously.

Most works focused on community tracking agree on the set of simple actions that involve entities of a dynamic network: node/edge appearance and vanishing. 
Indeed, such local and atomic operations can generate perturbations of the network topology able to affect the results produced by CD algorithms.
As a consequence of this set of actions, given a community $C$ observed at different moments in time, it is mandatory to characterize the transformations it undergoes. 
A first, formal, categorization of the transformations that involve communities were introduced in \cite{Palla2007}, which listed six of them (Birth, Death, Growth, Contraction, Merge, Split). 
A seventh operation, ``Continue", is sometimes added to these. 
In \cite{Cazabet2014}, an eighth operation was proposed (\emph{Resurgence}). 

\begin{figure*}
\centering
\includegraphics[width=.8\linewidth]{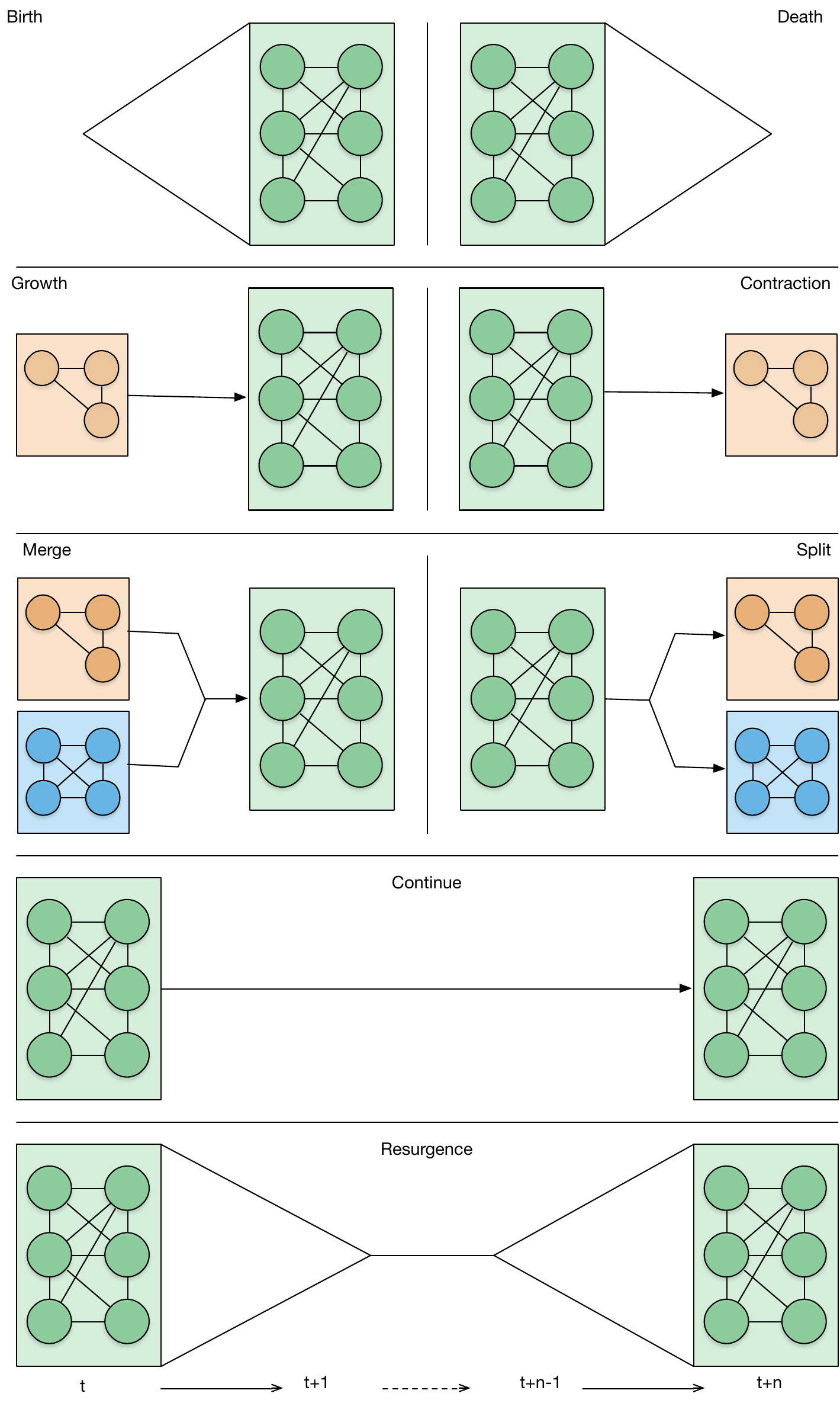}
\caption[Community transformation]{Community events. This toy example captures the eight events that regulates dynamic community life. In the first row Birth and Death; in the second row Growth and Contraction; in the third row Merge and Split; in the fourth row Continue; in the last row Resurgence.}
\label{img:events}
\end{figure*}

These operations, illustrated in Figure \ref{img:events}, are the following:
\begin{itemize}
    \item \emph{Birth}: the first appearance of a new community composed of any number of nodes;
    \item \emph{Death}: the vanishing of a community: all nodes belonging to the vanished community lose this membership;
    \item \emph{Growth}: new nodes increase the size of a community;
    \item \emph{Contraction}: some nodes are rejected by a community thus reducing its size;
    \item \emph{Merge}: two communities or more merge into a single one;
    \item \emph{Split}: a community, as consequence of node/edge vanishing, splits into two or more components;
    \item \emph{Continue}: a community remains unchanged;
    \item \emph{Resurgence}: a community vanishes for a period, then comes back without perturbations as if it has never stopped existing. 
    This event can be seen as a \empty{fake} death-birth pair involving the same node set over a lagged time period (example: seasonal behaviors). 
\end{itemize}

Not all operations are necessarily handled by a generic DCD algorithm. 
Even though the abstract semantics of such transformations are clear, different works propose to handle them differently. 
Among them, \emph{Merge}, \emph{Split} and \emph{Resurgence} are often defined with the support of ad-hoc similarity and thresholding functions as well as community transformation strategies. 
For instance, when two communities are chosen to be merged, several strategies can be followed to handle the \emph{Merge} action \cite{Cazabet2014}:
\begin{itemize}
    \item \emph{Absorption}: after the \emph{Merge}, one community disappears while the other persists. 
    In this scenario, defining a policy to decide which community will cease to exist is an open issue. 
    Among possible choices, we can cite:
    (i) Remove the oldest/youngest one;
    (ii) Remove the biggest/smallest one;
    (iii) Remove the community with the highest percentage of nodes included in the other.
    \item \emph{Replacement}: after the \emph{Merge} action, affected communities vanish, and a wholly new one forms. 
    This solution, although straightforward, causes sudden disruptions on the continuity of the evolution of communities.
\end{itemize}

Such complex scenario symmetrically applies to the specific implementation of all the remaining community actions. 

\begin{figure*}[ht]
\centering
\includegraphics[width=1\linewidth]{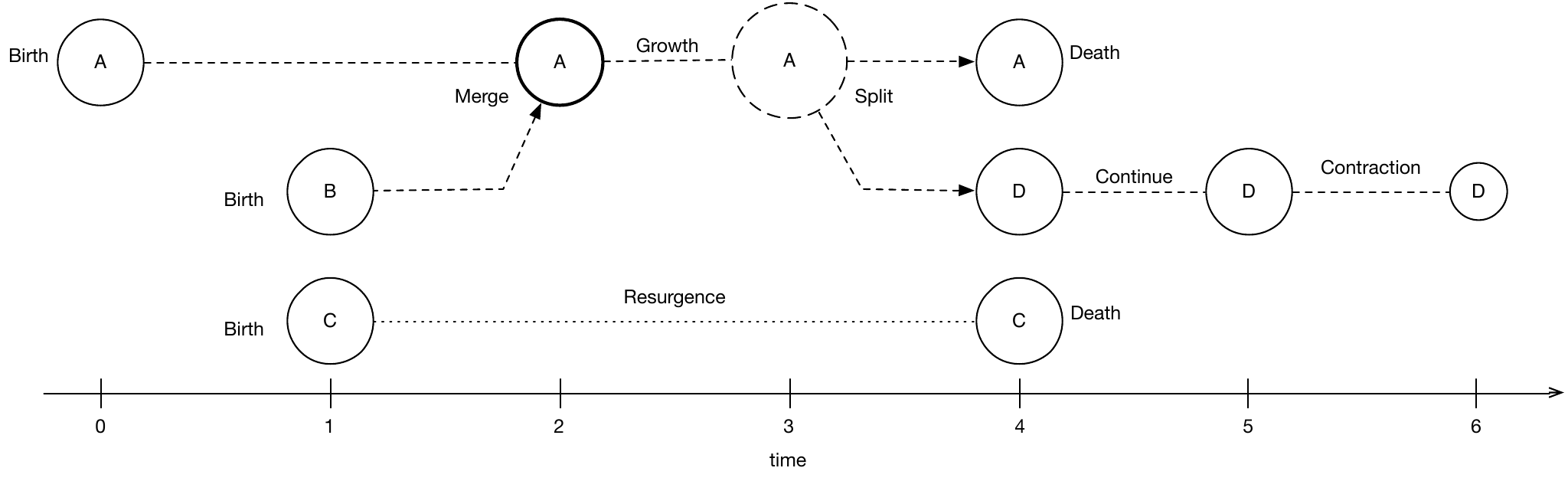}
\caption[Community Lifecyce]{Community Life-cycle. As time goes by, dynamic communities experience mutations that are often abstracted by adopting a predefined set of operations. This toy example highlights life-cycles of communities $A$ and $C$, composed of all the operations that modify a community structure, namely: birth, continue, growth, split, merge, contraction, resurgence, death.}
\label{img:lifecycle}
\end{figure*}

Despite the peculiarities introduced by each approach when defining topology transformations, the identified events allow to describe for each community its so-called \emph{life-cycle} (an example shown in Figure \ref{img:lifecycle}):

\begin{definition}[Community Life-cycle]
Given a community $C$, its community life-cycle (which univocally identifies $C$'s complete evolutive history) is composed of the Directed Acyclic Graph (DAG) such that: (i) the roots are \emph{Birth} events of $C$, and of its potential predecessors if $C$ has been implicated in \emph{Merge} events;
(ii) the leafs are \emph{Death} events, corresponding to deaths of $C$ and of its successors, if $C$ has been implicated in \emph{Split} events;
(iii) the central nodes are the remaining actions of $C$, its successors, and predecessors.
The edges of the tree represent transitions between subsequent actions in $C$ life.
\end{definition}

In principle, it is possible to perform a life-cycle analysis starting from the output of any generic dynamic community discovery approach.  

\subsection{Community Instability \& Temporal Smoothing}
\label{sec:instability}

One of the main issues encountered by dynamic community detection approaches is the instability of solutions. 
Such problem comes from the very nature of communities.
 
It is widely accepted that there is not a single valid decomposition in communities of a complex network issued from field data but, instead, several possible ones. 
Moreover, most algorithms generate partitionings in which a node is assigned unambiguously to one and only one community.
Indeed, we are aware that such scenario represents a simplification: nodes often belong to several communities, and the belonging to each of them is not necessarily binary (as illustrated by fuzzy partitions approaches).
Another issue is that the choice between one partition and another is somewhat arbitrary: moreover, a generic algorithm executed on the same network that experienced a few topological variations -- or even none in case of stochastic algorithms -- might lead to different results. 
To take a concrete example, if the problem of finding the solution of optimal modularity in a network is so complex, it is because many local maxima exist, whose modularity values are very close from the optimal one. 
Greedy approaches such as the Louvain method \cite{blondel2008fast} can find a local maximum; however, a slight variation of the initial network might switch from one local maximum to another, yielding significantly different partitions, both comparable in term of modularity score.

All such observations naturally lead to the general problem of identifying stable network decompositions. 
This is a major problem for dynamic community detection, as one cannot know if the differences observed between communities found on the network at time $t$ and those found at $t+1$ are due to the evolution of communities or to the algorithm's instability.

This problem has been explored in \cite{aynaud2010static}. 
The authors picked three well-known static community detection algorithms (WalkTrap \cite{pons2005computing}, Louvain \cite{blondel2008fast}, and Fast Greedy modularity \cite{clauset2004finding}), and evaluated the consistency of their results on slight modifications of a network. 
To do so, they compare the modularity \cite{newman2004fast,newman2004finding} and Normalized Mutual Information (NMI \cite{NMI}) scores between successive results, and quantify the number of transformations between two consecutive runs (where a \emph{transformation} is the change in affiliation of a node). 
The results revealed the extent of the problem: on a network composed of 9377 nodes and 24107 edges, constructed from a co-authorship network, the removal of a single node leads to NMIs between successive results of -- respectively to each algorithm-- 0.99, 0.9 and 0.75. 
The result is even more striking when looking at the number of unitary change: a single node modification leads respectively to approximately 500, 2000 and 3000 unitary change in communities.

A variety of solutions has been proposed to solve, or at least mitigate, this instability problem. 
Their common goal is to smooth-out the evolution of communities.
Among them, we can identify the following main categories: smoothing by bootstrap, explicit smoothing, implicit smoothing and global smoothing.
\\ \ \\
\noindent{\bf Smoothing by bootstrap.}
This technique (used in particular in \cite{Hopcroft2004,rosvall2010mapping}) consists of running multiple times the same algorithm on the same static network, searching for invariants, stable parts of the communities. 
The stable community cores, less unstable than raw community partitions, can be more efficiently tracked.
\\ \ \\ 
\noindent{\bf Explicit smoothing.}
Some algorithms (e.g., \cite{Lin2009,Folino2014}) explicitly introduce smoothing in their definition, requiring the partition found at step $t$ to have a certain degree of similarity with the partition found at $t-1$. 
These methods typically introduce a parameter $\alpha \in [0,1]$, that determines the trade-off between a solution that is optimal at $t$ and a solution maximizing the similarity with the result at $t-1$. 
\\ \ \\
\noindent{\bf Implicit smoothing.}
Some methods do not explicitly integrate the similarity between consecutive partitions but favor this similarity by construction. 
We can distinguish between two main scenarios to do so:
\begin{enumerate}
    \item Methods that optimize a global metric, such as the modularity, can use the communities found at the previous step as seeds for the current one (e.g., \cite{Shang2012,Gorke2010}). 
    The probability of finding similar solutions from step to step is increased by choosing an adequate algorithm, that will explore in priority good solutions around seeds.
    \item Other approaches (e.g., \cite{Cazabet2010,Xie2013,RPPG15}) do not try to optimize a global metric but update locally communities that have been directly affected by modifications between the previous and the current step. 
    In this case, the solution at step $t-1$ is kept unchanged, but for communities directly affected by changes in the network. 
    In a slowly evolving graph, this guarantees that most of the partition stays identical between two evolution steps.
\end{enumerate}

\noindent{\bf Global smoothing.}
Instead of smoothing communities at a given step by considering the previous one, algorithms (e.g., \cite{Aynaud2011,Mucha2010}) can search for communities that are coherent in time by examining all steps of evolution simultaneously. 
Such strategy can be applied, for instance, by creating a single network from different snapshots by adding links between nodes belonging to different snapshots and then running a community detection on this network. 
Another way to pursue in this rationale is to optimize a global metric on all snapshots simultaneously. 
In both cases, the problem of instability naturally disappears, as the detection of dynamic communities is performed in a single step.
\\ \ \\
When designing a DCD algorithm, the decision to adopt a specific smoothing strategy has non-negligible impacts: indeed, it can leads to computational constraints, narrows the applicative scenarios as well as impacts results it will be able to provide.
For all those reasons, in Section \ref{sec:classification} we describe a novel two-level taxonomy that exploits a characteristic of smoothing to discriminate among different classes of dynamic community discovery approaches.

\section{Classification}
\label{sec:classification}

\begin{figure*}[!th]
\centering
\includegraphics[width=1\textwidth]{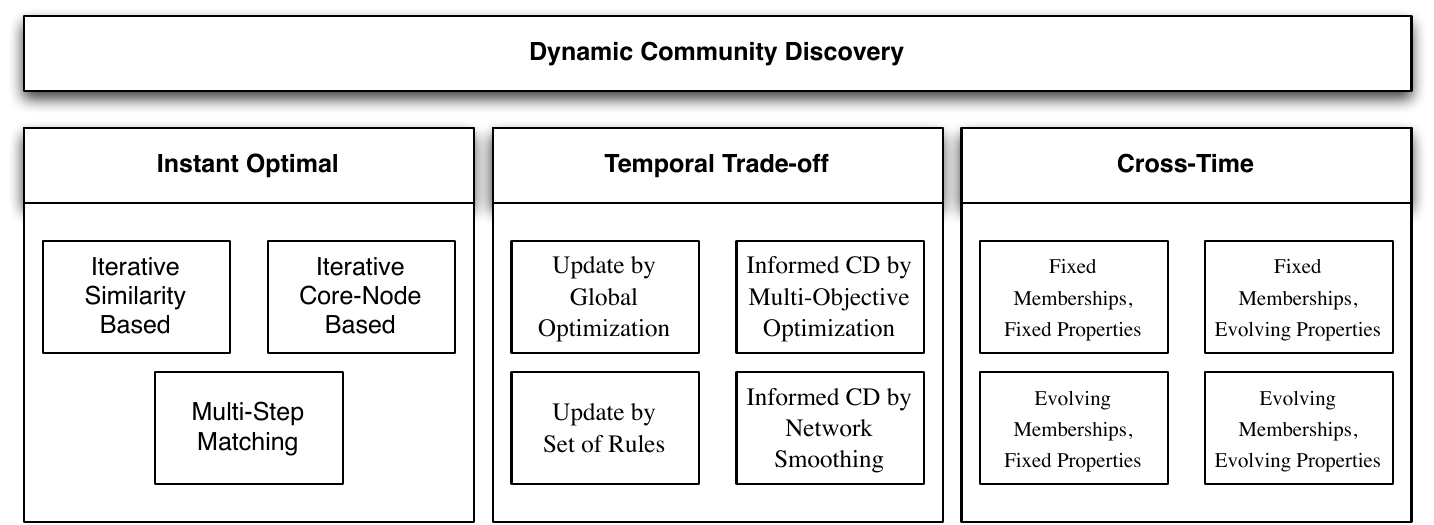}
\caption[Algorithm Taxonomy]{Proposed classification for DCD algorithms. There are three high-level classes, corresponding to different dynamic community definitions. Each of these classes is divided into subcategories, corresponding to different manners to solve a similar problem.}
\label{img:classification}
\end{figure*}

In this Section, we propose a classification, summarized in Figure \ref{img:classification}, of Dynamic Community Discovery methods. 

Three previous works, to the best of our knowledge, have proposed a classification of DCD algorithms, although considering much fewer methods that the current survey. 
In \cite{Aynaud2013}, identified classes are:
\begin{itemize}
    \item \textit{Two-Stage approaches}, methods that first detect clusters at each time step and then match them across different time-steps;
    \item \textit{Evolutionary Clustering}, methods that detect communities at time $t$ based on the topology of the graph at $t$ and on previously found community structures;
    \item \textit{Coupling graphs}, algorithms that first construct a single graph by adding edges between instances of nodes in different time-steps (coupling graph), and then run a classic CD on this graph.
\end{itemize}
Two-Stage approaches are then further divided into \textit{Core based}, \textit{Union Graph based} and \textit{Survival Graph based} methods, corresponding to different solutions to solve the problem of graph matching.

In \cite{hartmann2014clustering}, two high level categories are identified, \textit{Online} and \textit{Offline} methods. 
The survey focuses only on \textit{Online} algorithms, defined as methods that explicitly exploit information about the graph structure and the community structure of previous time steps, as opposed to \textit{Offline} ones that also use information from following time steps. 
Online methods are divided in:
\begin{itemize}
    \item \textit{Temporal Smoothness} methods that run a CD process from scratch at each step;
    \item \textit{Dynamic update} methods that do not run a process from scratch but instead update the methods found in previous time steps.
\end{itemize}

These first two classifications are complementary, and somewhat overlapping: \textit{Coupling graphs} methods are \textit{Offline}, but some \textit{Two-Stage} approaches can also be \textit{Offline}, since they match communities across all time-steps. 
\textit{Two-stage} approaches are a special case of \textit{Temporal Smoothness} without memory, while \textit{Evolutionary Clustering} methods encompass both \textit{Temporal Smoothness} and \textit{Dynamic update} methods.

Finally, in \cite{masuda2016guide}, the authors identify two different typologies of community discovery approaches in temporal networks:
\begin{itemize}
    \item Approaches focused on determining \textit{Community evolution} that, with certain statistical accuracy, aim to detect when and how reorganisations of communities take place. In this family fall the \textit{two-stage} as well as \emph{model-based} approaches.
    \item \textit{Single community structure} methods whose goal is to find a single partition of the dynamic graph by considering time as a third dimension of the data, added to the classical two expressed by the adjacency matrix. Among them fall the \textit{Tensor factorization} approaches like the one introduced in \cite{gauvin2014detecting}, where a single partition is identified and the \emph{strength} of each community per temporal snapshot is computed.
\end{itemize}

Some methods we found in the literature were not belonging to any of these categories or could be associated with several of them.

We decided to shape our classification as a two-level one, building it upon concepts shared by existing classifications.

In our taxonomy, the higher level identifies three different definitions of what are Dynamic Communities, without assumptions on the techniques used to find them. 
These high-level classes are then refined into subcategories, which correspond to different techniques used to find communities corresponding to this definition.

There are three classes of approaches at the higher level of classification:
\begin{itemize}
    \item The first one (\textit{Instant-optimal CD}) considers that communities existing at $t$ only depend on the \textbf{current} state of the network at $t$. Matching communities found at different steps might involve looking at communities found in previous steps, or considering all steps, but communities found at $t$ are considered optimal, w.r.t. the topology of the network at $t$.  Approaches falling in this class are \textbf{non-temporally smoothed}.
    \item In the second class (\textit{Temporal Trade-off CD}), communities defined at an instant $t$ do not only depend on the topology of the network at that time, but also on the \textbf{past} evolutions of the topology, \textbf{past} partitions found, or both. Communities at $t$ are therefore defined as a trade-off between optimal solution at $t$ and known past. They do not depend on future modification, an important point for ``on the fly" CD. Conversely, from the approaches falling in the previous class, Temporal Trade-off ones are \textbf{incrementally temporally smoothed}.
    \item In the third class (\textit{Cross-Time CD}), the focus shifts from searching communities relevant at a particular time to searching communities relevant when considering the whole network evolution. Methods of this class search a single partition directly for all time steps. Communities found at $t$ depends \textbf{both on past and future} evolutions. Methods in this class produce communities that are \textbf{completely temporally smoothed}.

\end{itemize}
For each of these classes, we define subcategories, corresponding to different techniques used to solve a related problem. 
In the following we address each one of them, discussing their advantages and drawbacks. As we will highlight in the following, each class -- due to its definition -- is likely to group together approaches designed to work on either Network Snapshots (SN) or Temporal Networks (TN).

Please note that to keep our classification concise, we do not include the description of each method here. 
We provide a short description of methods included in each category in the \textbf{Appendix}.

\subsection{Instant-optimal Communities Discovery}
\label{subsec:instantoptimal}

\begin{figure*}[!t]
\centering
\includegraphics[width=.9\textwidth]{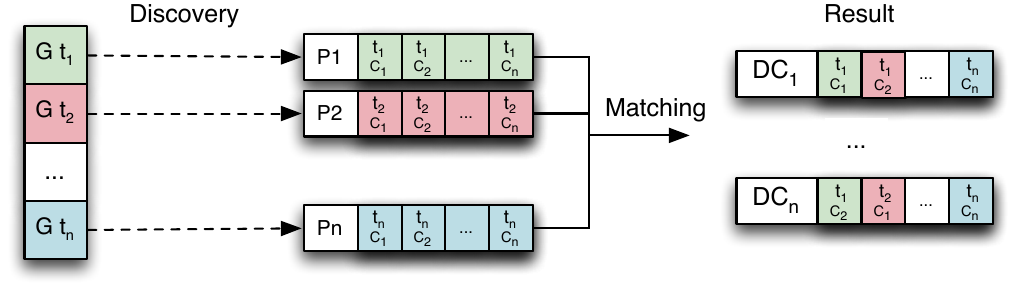}
\caption[Instant Optimal]{Instant Optimal. The identification of dynamic communities is conceptually decomposed in two subproblems: (i) independent extraction of the network partition for each time step, and (ii) matching of communities belonging to distinct network observations. Legend: G: Graph, P: Partition, DC: Dynamic Community, C: Community}
\label{img:instantoptimal}
\end{figure*}

This first class of approaches is derived directly from the application of static community discovery methods to the dynamic case. 
A succession of steps is used to model network evolution, and for each of them is identified an optimal partition, as shown in Figure \ref{img:instantoptimal}. 
Dynamic communities are defined from these optimal partitions by specifying relations that connect topologies found in different, possibly consecutive, instants.

A typical Instant-optimal technique has been identified as \textit{Two-Stage Approach} in the literature \cite{Aynaud2013}. 
It can be described as a two steps process:
\begin{enumerate}
  \item \emph{Identify}: detect static communities on each step of evolution;
  \item \emph{Match}: align the communities found at step $t$ with the ones found at step $t-1$, for each step.
\end{enumerate}
It can be noted however that \textit{Two-Stage Approaches} and \textit{Instant-optimal} communities do not encompass the same methods: indeed, two-stage techniques can be used with communities smoothed-out using previous communities, and therefore not Instant-optimal. 
On the contrary, some methods find instant-optimal communities but match these communities considering all partitions simultaneously, instead of using an iterative process.
In the following, we discuss the advantages and drawbacks of this class of algorithms, and the dynamic network models it can be applied to.
\\ \ \\
\noindent{\bf Advantages}:
The major benefit of adopting this approach is that it is built directly on top of existing works on static community detection. 
Since communities are Instant-Optimal, usual DC algorithms can be used at each step. 
The matching process can also be derived from existing literature since set matching is an extensively studied problem. 
Another advantage of this method is that it easily enables parallelization. 
Community detection can be a time-consuming process on large networks, and since the communities at each step are found independently, it is possible to run the detection at each step in parallel.
\\ \ \\
\noindent {\bf Drawbacks}:
The main limit of this approach is the instability of community detection algorithms (see Section \ref{sec:instability} for more details on instability). Since the same algorithm can find different communities on the same network, it is hazardous to distinguish between changes due to the evolution of the community structure and changes due to the instability of algorithms. 
Recent approaches propose solutions to circumvent this weakness, by studying the most stable part of communities, called \textit{community cores} \cite{rosvall2010mapping}, or by considering all partitions simultaneously, instead of matching only with earlier ones \cite{Goldberg2011}. 
\\ \ \\
\noindent{\bf Network models}:
Due to its definition, algorithms that fall in this class can be used only with Network Snapshots, not with Temporal Networks. 
As the community detection needs to be performed entirely from scratch at each evolution step, the whole network needs to be considered, not only the changes between one step and another.
\\ \ \\
\noindent{\bf Subcategories}:
The first step of the process, identifying communities relevant at each time step, can usually be done with any CD algorithm. 
However, the methods used to match communities found at different steps differ. We identified three subclasses: (i) \emph{Iterative similarity-based}  (ii) \emph{Iterative core-based} (iii) \emph{Multi-step} matching.

\subsubsection{Iterative similarity-based approaches}  
In similarity-based approaches, a quality function is used to quantify the similarity between communities in different snapshots (e.g., Jaccard among node sets). 
Communities in adjacent time-steps having the highest similarity are considered part of the same dynamic community. Particular cases can be handled, such as a lower threshold of similarity, or the handling of $1$ to $n$ or $n$ to $n$ matchings, thus defining complex operations such as split and merge. 

Methods in this subcategory: \cite{Hopcroft2004,Bourqui2009,Palla2007,greene2010tracking,rosvall2010mapping,Takaffoli2011,BKP11,Brodka2013,Dhouioui2014,NS15}. 

Refer to \textbf{Appendix \ref{subcat1}} for methods description.

\subsubsection{Iterative Core-nodes based approaches}
Methods in this subcategory identify one or several special node(s) for each community -- for instance those with the highest value of a centrality measure -- called \textbf{core-nodes}. Two communities in adjacent time-steps containing the same core-node can subsequently be identified as part of the same dynamic community. It is thus possible to identify splits or merge operations by using several core-nodes for a single community. 

Methods in this subcategory: \cite{Wang2008,Chen2010}

Refer to \textbf{Appendix \ref{subcat2}} for methods description.

\subsubsection{Multi-step matching}
Methods in this subcategory do not match communities only between adjacent time-steps. Conversely, they try to match them even in snapshots potentially far apart in the network evolution. The matching itself can be done, as in other sub-categories, using a similarity measure or core-nodes. The removal of the adjacency constraint allows to identify \textit{resurgence} operations, i.e. a community existing at $t$ and $t+n$ but not between $t+1$ and $t+(n-1)$. It can, however, increase the complexity of the matching algorithm, and \textit{on-the-fly} detection becomes inconsistent, as past affiliations can change.

Methods in this subcategory: \cite{Falkowski2006,FS07,Goldberg2011,morini2017revealing}

Refer to \textbf{Appendix \ref{subcat3}} for methods description.

\subsection{Temporal Trade-off  Communities Discovery}
\label{sec:inf_iter}
\begin{figure*}[!t]
\centering
\includegraphics[width=.9\textwidth]{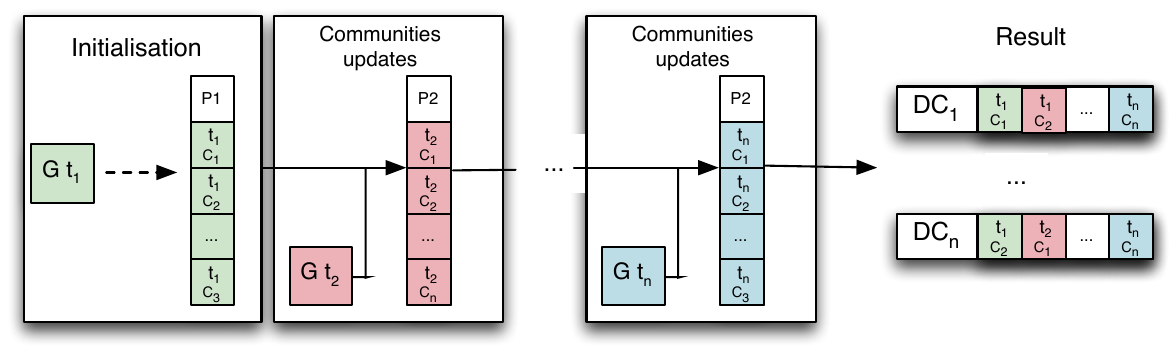}
\caption[Temporal Trade-off]{Temporal Trade-off. After a \emph{bootstrap}  phase, communities are found iteratively at each step by exploiting information gathered from previous networks statuses and partitions. Legend: G: Graph, P: Partition, DC: Dynamic Community, C: Community}
\label{img:temporaltradeoff}
\end{figure*}

Algorithms belonging to the \emph{Temporal Trade-off} class process iteratively the evolution of the network. 
Moreover, unlike Instant optimal approaches, they take into account the network and the communities found in the previous step -- or $n$-previous steps -- to identify communities in the current one.
DCD algorithms falling into this category can be described by an iterative process:
\begin{enumerate}
    \item \emph{Initialization}: find communities for the initial state of the network;
    \item \emph{Update}: for each incoming step, find communities at step $t$ using graph at $t$ and past information.
\end{enumerate}
As for the \emph{Instant-optimal} class, this template, exemplified in Figure \ref{img:temporaltradeoff}, can be implemented by several typologies of approaches. 
In the following we discuss its advantages and drawbacks, and the dynamic network models it can be applied to.
\\ \ \\
\noindent{\bf Advantages}:

\emph{Temporal Trade-off} approaches allow to cope with the instability problem that affects \emph{Instant-optimal} ones (see Section \ref{sec:instability}), while not diverging much from 
 the usual community detection definition (a partition is searched-for at each step).
Most methods of this class take advantage of the existing community structure at step $t-1$ to speed up the community detection at $t$.
\\ \ \\
\noindent{\bf Drawbacks}:
In the general case, it is not possible to easily parallelize the community detection, as each step needs the communities found at the previous ones as input. 
Another potential issue is the long-term coherence of dynamic communities. 
Since at each iteration the identified partition depends on previous ones, \emph{Temporal Trade-off} approaches are subject to the risk of an avalanche effect: communities can experience substantial drifts compared to what a static algorithm would find on the static network corresponding to the state of the system at a given time.
\\ \ \\
\noindent{\bf Network Models}:
This algorithmic schema has been used with both dynamic networks representations introduced in Section \ref{subsec:rep}. 
When using Temporal Networks, only the incremental modifications of the network are considered to make the communities evolve, usually using local rules. 
Conversely, when using Network Snapshots, a process is usually run on the whole graph, as for a static algorithm, but taking into account past information.
\\ \ \\
\noindent{\bf Subcategories}:
The \emph{Temporal Trade-off} template can be satisfied by very different algorithmic approaches. 
Among them, we identified four subcategories.

The first two, namely \emph{Update by Global Optimization} and \emph{Update by a set of rules}, consist in \textbf{updating} the partition found at the previous step, using global or local approaches. 

The last two, namely \emph{Informed CD by Multi-objective Optimization} and \emph{Informed CD Network Smoothing}, run a community detection from scratch for each snapshot while considering previous steps information. 

\subsubsection{Update by Global Optimization}
A popular method for static CD is to optimize a Global Quality Function, such as Modularity or Conductance. 

This process is typically done using a heuristic (gradient descent, simulated annealing, etc\dots) that iteratively merges and/or splits communities, and/or moves nodes from a community to another until a local or global maximum is reached. 

Methods in this subcategory use the partition existing at $t$ as a seed to initialize a Global optimization process at $t+1$. 
The heuristic used can be the same as in the static case or a different one. 
A typical example of a method in this category \cite{aynaud2010static} consists in using the Louvain algorithm: in its original definition, at initialization, each node is in its community. 
The \textit{partition update} version of it consists in initializing Louvain with communities found in the previous step.

Methods in this subcategory: \cite{ME09,Gorke2010,aynaud2010static,Bansal2011,Shang2012,AHS14}

Refer to \textbf{Appendix \ref{subcat4}} for methods description.

\subsubsection{Update by Set of Rules}
Methods in this subcategory consider the list of network changes (edge/node apparitions, vanishing) that occurred between the previous step and the current one, and define a list of rules that determine how networks changes lead to communities update. 
Methods that follow such rationale can vary significantly from one to the other.

Methods in this subcategory: \cite{Falkowski2008,Nguyen2011,Cazabet2010,Nguyen2011b,cazabet2011simulate,Agarwal2012,Duan2012,Gorke2012,Huang2013,Xie2013,Lee2014,Zakrzewska2015,RPPG15}

Refer to \textbf{Appendix \ref{subcat5}} for methods description.

\subsubsection{Informed CD by Multi-objective optimization} 
When addressing community detection on Snapshot Graphs, two different aspects need to be evaluated for each snapshot: partition quality and temporal partition coherence. 
The multi-objective optimization subclass groups together algorithms that try to balance both of them at the same time so that a partition identified at time $t$ represents the natural evolution of the one identified at time $t-1$. 
This approach optimizes a quality function of the form: 
\begin{equation}
    c = \alpha CS + (1-\alpha) CT
\end{equation}
with $CS$ the cost associated with current snapshot (i.e. how well the community structure fits the graph at time $t$), and $CT$ the smoothness w.r.t. the past history (i.e., how different is the actual community structure w.r.t.  the one at time $t-1$) and $\alpha \in [0,1]$ is a correction factor. 

An instantiation of such optimization schema could be defined using \emph{modularity} \cite{newman2004finding} as $CS$ and \emph{Normalized Mutual Information} \cite{NMI} as $CT$ as done in \cite{Folino2010}.\\

Methods in this subcategory: \cite{Zhou2007,Tang2008,Lin2009,Lin2008,Yang2009,Folino2010,Sun2010,GJMZ12,Kawadia2012,CD15,Gorke2013}

Refer to \textbf{Appendix \ref{subcat6}} for methods description.

\subsubsection{Informed CD by network smoothing}
Methods in this subcategory search for communities at $t$ by running a CD algorithm, not on the graph as it is at $t$, but on a version of it that is \textit{smoothed} according to the past evolution of the network, for instance by adding weights to keep track of edges' age. In a later step, communities are usually matched between snapshots, as in typical \textit{Two-Stage} approaches.
Contrary to previous subcategories, it is not the previous communities that are used to take the past into account, but the previous state of the network.

Methods in this subcategory: \cite{Kim2009,Guo2014,Xu2013b}

Refer to \textbf{Appendix \ref{subcat7}} for methods description.

\subsection{Cross-Time Communities Discovery}
\label{subsec:crosstime}
\begin{figure}[!t]
\centering
\includegraphics[width=.6\columnwidth]{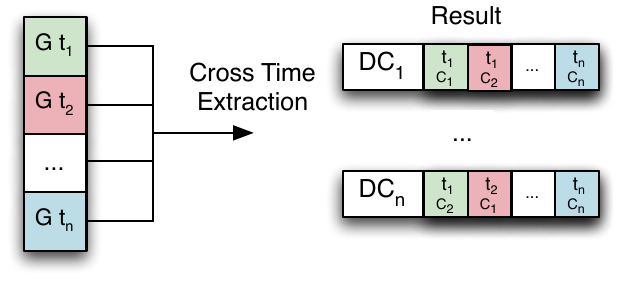}
\caption[Cross Time]{Cross Time. Communities are extracted by leveraging, all at once, the information gathered by the complete temporally annotated history of the analyzed network. Legend: G: Graph, P: Partition, DC: Dynamic Community, C: Community}
\label{img:crosstime}
\end{figure}
Approaches belonging to the \emph{Cross-Time} class do not consider independently the different steps of the network evolution. 
On the contrary, DCD is done in a single process, considering -- at the same time -- all states of the network. 
A visual representation of this class of approaches is shown in Figure \ref{img:crosstime}. 

An example of algorithmic procedures that fall in this class is the following:
\begin{enumerate}
    \item \emph{Network Transformation}: from a sequence of snapshots, a single network is created in which each meta-node corresponds to the presence of a node at a given snapshot.
    Two kinds of edges are defined on this network: (i) the usual relationship between nodes at the same time-step, and (ii) additional edges that link nodes belonging to different, adjacent, time-steps.
    \item \emph{Community Detection}:
    A static community detection algorithm is run on this transversal network: identified communities contain nodes belonging to different time steps, interpretable as dynamic communities.
\end{enumerate} 
\noindent{\bf Advantages}
This class of algorithms does not suffer from problems of instability and community drift that affect previous ones. 

In theory, these methods have more potential to deal with local anomalies and slow evolution running through multiple steps, that iterative processes might not be able to perceive, due to their narrower focus.
\\ \ \\
\noindent{\bf Drawbacks}
\emph{Cross-Time} approaches, unlike previous ones, are not based on the usual principle of a unique partition being associated with each step of the graph evolution process. 
It, therefore, requires to develop novel techniques and to make new assumptions on the nature of dynamic communities. 
As a result, most methods suffer from constraints such as fix number of communities and absence of operations such as merge or split.

Another drawback is that such approaches are not able to handle on-the-fly/real-time community detection. 
Indeed, since their computation needs a complete knowledge of all the network history, in the presence of a new evolutive step the partition extraction needs to be performed from scratch. 
\\ \ \\
\noindent{\bf Network Models}
Both Network Snapshots and Temporal Networks can be used by Cross-Time approaches.
\\ \ \\

\noindent{\bf Subcategories}:
All algorithms in the \emph{Cross-Time} class do not have the same constraints on the nature of searched communities. 
We can classify them in four categories: (i) \emph{Fixed Memberships, fixed properties} (ii) \emph{Fixed memberships, evolving properties} (iii) \emph{Evolving memberships, fixed properties} (iiii) \emph{Evolving memberships, Evolving properties}.

\subsubsection{Fixed Memberships, fixed properties}
Methods in this subcategory do not allow nodes to switch communities, nor communities to appear or disappear. 
They also assume that communities stay the same throughout the studied period. 
As a consequence, they search for the best partition, \textit{on average}, over a time period. 
In most cases, they improve the solution by slicing the evolution in several time periods, each of them considered homogeneous, and separated by dramatic changes in the structure of the network, a problem related to change point detection \cite{peel2015detecting}.

Methods in this subcategory: \cite{Sun2007,Duan2009,Aynaud2011}

Refer to \textbf{Appendix \ref{subcat8}} for methods description.

\subsubsection{Fixed memberships, evolving properties}
Methods in this subcategory are aware that communities are not homogeneous along time. 
For instance, nodes in a community can interact more actively during some recurrent time-periods (hours, days, weeks, etc.), or their activity might increase or decrease along time. 
To each community -- whose membership cannot change -- they assign a temporal profile corresponding to the evolution of its activity.
 
Methods in this subcategory: \cite{gauvin2014detecting,matias2018semiparametric}

Refer to \textbf{Appendix \ref{subcat9}} for methods description.

\subsubsection{Evolving memberships, fixed properties}
Methods in this subcategory allow node memberships to change along time, i.e., nodes can switch between communities. 
However, because they use Stochastic Block Models approaches, for which the co-evolution of memberships and properties of communities is not possible (see \cite{matias2016}), the number of communities and their density is fixed for the whole period of analysis.

Methods in this subcategory: \cite{Yang2009,Yang2010,ishiguro2010dynamic,herlau2013modeling,matias2016,ghasemian2016detectability}

Refer to \textbf{Appendix \ref{subcat10}} for methods description.

\subsubsection{Evolving memberships, evolving properties}
Methods in this category do not impose constraints on how dynamic communities can evolve: nodes can switch between them, communities can appear or disappear, and their density can change. 
Three main approaches are currently present in this category: 
\begin{itemize}
    \item In \cite{JRF07} and \cite{Mucha2010}, edges are added between nodes in different snapshots. Static community detection algorithms are consequently run on this trans-temporal network
    \item In \cite{VLM15} and \cite{HMNS16}, the authors search for persistent cliques of a minimal size in link streams
    \item In \cite{xu2014dynamic}, a dynamic stochastic block model is defined that allows both  affiliations and densities to change. 
\end{itemize}

Methods in this subcategory: \cite{JRF07,Mucha2010,VLM15,HMNS16,xu2014dynamic}

Refer to \textbf{Appendix \ref{subcat11}} for methods description.

\subsection{Discussion}
As we have seen in this section, all three classes of approaches have advantages and drawbacks; none is superior to the other. 
Nevertheless, we have observed how each one of them is more suitable for some use cases. 

For instance, if the final goal is to provide on-the-fly community detection on a network that will evolve in the future, approaches that fall in the first two classes represent the most suitable fit. 
If the analytical context requires working with a fine temporal granularity, therefore modeling the observed phenomena with temporal networks, it is strongly suggested to avoid methods of the first class, that can only deal with snapshots.

The first layer of our taxonomy can thus be used to provide guidance and recommendations on which approach (or class of approaches) select given a specific problem formulation. 
For instance, we can observe how, 
\begin{itemize}
    \item \emph{Instant Optimal} approaches are the best choice when the final goal is to provide communities which are as good as possible at each step of the evolution of the network.
    \item \emph{Cross-Time} approaches are the best choice when the final goal is to provide communities that are coherent in time, in particular over the long-term.
    \item \emph{Temporal Trade-off} approaches represent a tradeoff between these other two classes: they are the best choice in case of continuous monitoring, rapidly evolving data, and in some cases limited memory applications.
\end{itemize} 

The second layer of classification groups together approaches that share the same rationale. 
This further classification is useful, in our opinion, to those researchers that need to frame their method within a specific literature.
Identifying the particular family a DCD approach belongs to is valuable to understand which are its real competitors and, doing so, to better organize comparative analysis, reducing the bias introduced by the slightly different problems addressed (as well as community definitions adopted) by alternative methods.

\section{Evaluation}
\label{sec:evaluation}
So far we introduced our taxonomy and framed the DCD algorithms we identified within it. 
In this section, we will address a very sensitive theme related to any data mining task: evaluation.
Indeed, finding a reliable way to evaluate partition quality is a significant issue to address while approaching community discovery.
One of the main issues of the community discovery lies in the absence of a unique definition of \emph{community}: each scholar follows its strategy to compare the results obtained by its algorithm with those produced by other \emph{state-of-the-art} methods, each one defining specific characteristics that a proper network partition should express.

Often, the comparison is made amongst methods having similar characteristics: algorithms that optimize the same quality function (i.e., modularity, conductance, density\dots), that produce overlapping/crisp/hierarchical network partitions, which are designed to work on directed/undirected graphs,\dots.
To shed some lights on the most popular evaluation strategies adopted in literature, in the following we categorize them and discuss testing environments used to evaluate community partitions.
In particular, in Section \ref{subsec:generators}, we describe how synthetic network generators are often used to design controlled experiments: there we introduce two classes of benchmarks built upon them, namely \emph{static} and \emph{dynamic} ones.
Finally, in Section \ref{subsec:ev_methods} we discuss two families of approaches designed to assess community quality, namely \emph{internal} and \emph{external} evaluation.

\subsection{Synthetic Network Generators}
\label{subsec:generators}

Several network properties can be used to characterize real-world phenomena: network modeling aims to replicate them thus allowing for the generation of synthetic datasets that, at least to some extent, can be used as analytical proxies.
The general aim of network modeling is to capture some essential properties lying behind real-world phenomena and replicate them while generating synthetic data, imposing only a few simple constraints. 

Several models of increasing complexity and realism were proposed during the 20th century, the most famous being Erd\"os and R\'enyi random graphs \cite{ER59}, Watts and Strogatz \emph{small-world networks} \cite{Watts-Colective-1998}, Barab\'asi and Albert \emph{Preferencial attachment} model \cite{barabasi1999emergence}, Leskovec's \emph{Forest Fire} \cite{LKF05} and \emph{Community-Affiliation Graph} \cite{yang2012structure}.
For a more detailed presentation of the history of network models, refer to \cite{rossetti2015social}

\bigbreak


Complex network modeling studies generated a new field of research: synthetic network generators.
Generators allow scientists to evaluate their algorithms on synthetic data whose characteristics resemble the ones that can be observed in real-world networks. 
The main reason behind the adoption of network generators while analyzing the performances of a community discovery algorithms is the ability to produce benchmark datasets that enable for:
\begin{itemize}
\item \emph{Controlled Environment testing:} Network generators allow to fine-tune network characteristics -- i.e.,  network size and density.  Such flexibility enables an extensive algorithm evaluation on networks having different characteristics but generated to follow similar topologies. 
Generators make possible, given a CD algorithm, to evaluate:
\begin{itemize}
    \item \emph{Stability:} the performance of a CD approach can be evaluated on a high number of network instances having similar properties to provide an estimate of the algorithmic stability;
    \item \emph{Scalability:} synthetic graphs can be used to test the actual scalability of an algorithm while increasing the network size.
\end{itemize}
\item \emph{Ground-truth testing:} some network generators provide as a by-product a ground-truth partition of the generated network. Such partition can be used to evaluate the one provided by the tested algorithm.
\end{itemize}
Two families of network generators have been described to provide benchmarks for community discovery algorithms: generators that produce \emph{static} graphs-partitions and generators that describes \emph{dynamic} graphs-partitions. 
Surprisingly, the formers are often used to evaluate also DCD algorithms: this happens because they are more widespread than the latter, and allows comparison with static algorithms. 
Of course, they can only be used to evaluate the quality of the detection at time $t$, and not the smoothness of communities.
\\ \ \\
\noindent{\bf Static Benchmarks.}
The classic and most famous static benchmarks are the Girvan-Newman  (introduced by Girvan and Newman \cite{girvan2002community}) and the LFR (introduced by Lancichinetti et al. \cite{NMI}).
The Girvan-Newman benchmark, also known as GN, is built upon the so-called \emph{planted l-partition} model described in \cite{condon2001algorithms,brandes2003experiments,gaertler2007significance} this generator takes as input a given ground truth partition (generated as as equally sized Erd\"os-Re\'nyi random graphs \cite{ER59}) and two parameters identifying respectively the probabilities of intra and inter-clusters links. 
\emph{Planted l-partition} models (also known as \emph{ad-hoc} models) where proposed to produce graph having different characteristics: overlapping community structures \cite{sawardecker2009detection,danon06} as well as weighted \cite{fan2007accuracy} and bipartite \cite{guimera2007module} graphs were modeled in order to better mimic real network characteristics.
To cope with the limitation of the GN benchmark (Poisson degree distribution, equal community sizes), and to fill the gap among Erd\"os-R\'enyi graphs and real ones, in \cite{NMI} was introduced .
The networks generated by this model have both node degrees and community sizes following a power law distribution. 
LFR  has also been generalized to handle weighted and directed graphs, as well as to generate overlapping communities \cite{lancichinetti2009benchmarks}. 
\\ \ \\
\noindent{\bf Dynamic Benchmarks.}
Recently a few papers have proposed extensions of GN and LFR that introduce topology dynamics as well as entirely new dynamic benchmarks.

In \cite{Lin2008}, a variant of the GN model is introduced to evaluate the FaceNet framework. There, the authors introduce network dynamics by generating different graph snapshots and interpolating from a snapshot to its subsequent by (i) randomly selecting a fixed number of nodes from each community (ii) removing them and (iii) allocating them to different communities.

In \cite{greene2010tracking}, a dynamic benchmark built upon LFR is introduced: starting from a static synthetic graph with ground-truth communities such benchmark randomly permuted 20\% of the node memberships to mimic node-community migrations.

In \cite{granell2015benchmark}, a new model based on Stochastic Block Models is proposed: a graph is split into $q$ subgraphs where nodes belonging to the same subgraph are connected with probability $p_{in}$, while edges connecting subgraphs have probability $p_{out}$ then ad-hoc procedure make the network evolve generating community events. 
In \cite{bazzi2016generative}, a generic approach is proposed to generate both multilayer and temporal networks with a community structure. The method adopts a 2 steps process: 1) Generating a multi-layer partition satisfying user-defined parameters 2) Sampling edges consistent with the partition. The underlying model of edges and partition is the degree-corrected SBM. An interlayer dependency tensor allows to define how strongly the community structure in one layer depends on other layers. In the case of dynamic networks, the layer corresponding to time $t$ typically depends only on the network at time $t-1$. The model does not support community events, but provide an implementation in MATLAB\footnote{\url{https://github.com/ MultilayerBenchmark/MultilayerBenchmark/}}.

Finally, in \cite{rossetti2017} RDyn, a new benchmark specifically tailored for the DCD problem, is introduced. 
RDyn\footnote{RDyn code available at: \url{https://goo.gl/WtLg4V}} generates dynamic network topologies (both in SN and TN format) as well as temporally evolving ground-truth communities.

\subsection{Methodologies}
\label{subsec:ev_methods}
Community partitions can be evaluated pursuing \emph{external} as well as \emph{internal} quality analysis strategies.
The former methodologies assume the existence of an external ground truth that needs to be retrieved or a specific partition quality score to optimize. 
The latter instead focus on the inspection and description of topologies identified as well as on the evaluation of the approach's complexity and scalability.\\

\noindent{\bf External: ground-truth communities.}
We have seen how Girvan-Newman and LFR were designed to produce tunable network structures having ground-truth partitions. 
Moreover, several real-world datasets also come with associated community annotations\footnote{See: \url{https://snap.stanford.edu/data/index.html#communities}}.

\begin{figure}[!t] 
\centering
\includegraphics[width=.6\columnwidth]{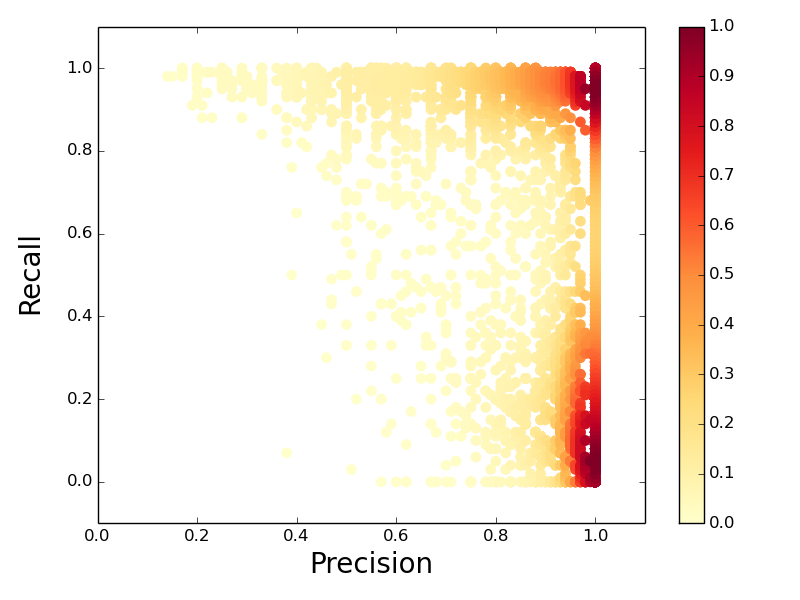} 
\caption{\emph{F1-community} measure scatter plot. Each point represent the (precision,recall) pairs for a given match: the deeper the color the more the communities sharing the same values of (precision,recall) values. The optimal matching value is the upper right corner (precision and recall  equals to 1): communities having high precision and low recall (bottom-rigth corner) underestimate the ground-truth ones, communities having low precision and high recall (top-left corner) overestimate the ground truth ones.
}
\label{img:f1_com} 
\end{figure}

Ground-truth community structures have well-known characteristics (i.e., fixed size, density, cut-ratio\dots) and/or semantic coherence (i.e., node label homophily on a given external attribute).
For these reasons, a way to evaluate the effectiveness of community discovery algorithms consists in comparing the divergence between the partition they produce and the planted one. 
Although several criticisms were opposed to this evaluation methodology \cite{peel2017ground} (e.g., since the community discovery is an ill-posed problem it is not assured that the planted partition is the optimal one for all quality functions optimized by existing algorithms), it is still the most widely spread to compare different algorithmic approaches. 
The common way to assess how a given partition resembles the ground-truth one is to compute the \emph{Normalized Mutual Information score} (NMI, \cite{NMI,NMIo,NMIoh}) a measure of similarity borrowed from information theory, defined as \cite{RPR16}: 
\begin{equation}
\label{eq:nmi}
NMI(X, Y)= \frac{H(X)+H(Y)-H(X,Y)}{(H(X)+H(Y))/2}
\end{equation}
where $H(X)$ is the entropy of the random variable $X$ associated to an identified community, $H(Y)$ is the entropy of the random variable $Y$ associated to a ground truth one, and $H(X, Y)$ is the joint entropy. 
NMI is defined in the interval [0,1] and is maximal when the compared communities are identical. 
One drawback of NMI is that assuming an approximate size $z$ for the compared community sets its computation requires $O(z^2)$ comparisons, a complexity that makes it often unusable to evaluate partitions of large-scale networks. 
To cope with the high computational complexity of such method in recent years, several approaches were proposed \cite{cazabet2015using,RPPG15,RPR16}.
As an example, in \cite{RPR16,RPPG15} the \emph{F1-community} score has been introduced. 
In such work the evaluation problem is solved as a classification task: 
\begin{itemize}
    \item network nodes are labeled according to their ground-truth community label;
    \item each community identified by the tested algorithm is matched to the ground-truth one whose label is shared by the majority of its nodes;
    \item precision and recall are computed for each community by considering the actual nodes and the real ones;
    \item the \emph{F1-community} score is computed as the average of the harmonic mean of precision and recall of the matched communities.
\end{itemize}
This approach (used in \cite{RPPG15,rossetti2017} where a normalized version, NF1\footnote{NF1 code available at: \url{http://goo.gl/kWIH2I}}, is proposed to cope with community overlap and redundancy) requires $O(z)$ to be computed (assuming an already labeled graph). 
Moreover, it allows to graphically visualize the performances of a given algorithm via density scatter plots. 
As an example, Figure \ref{img:f1_com} shows a visual representation of \emph{F1-community} where the communities identified by a CD method are compared with ground-truth ones.

Although real datasets with ground-truth communities as well as the LFR/GN benchmarks are commonly used to perform testing of DCD approaches by considering the network at a particular instant of its evolution, it is mandatory to underline one huge deficiency: such methods do not take into account the temporal dimension. 
In particular, GN and LFR were designed to allow ground-truth testing for classical static community discovery algorithms. 
For this reason, their use in conjunction with dynamic approaches raise several practical issues: how to generate subsequent (coherent) network snapshots? 
Is it possible to revise them to produce interaction networks? If so how can they reliably mimic edge/node vanishing?
One of the open problems that afflict DCD is, as of today, the evaluation of the produced communities: how can time-aware ground-truths partitions be defined (even in synthetic scenarios)? 
As discussed in Section \ref{subsec:generators}, so far only a few approaches have been designed to address these latter issues (e.g., RDyn \cite{rossetti2017}).
\\ \ \\
\noindent{\bf External: quality function.}
In absence of ground-truth communities, a common way to compare different algorithms is to rank their partitions w.r.t. a community quality score. 
The algorithm that produces the partition reaching the highest score is then, by definition, considered the best one for the analyzed network. 

\emph{Modularity} \cite{newman2004fast,newman2004finding} is probably the most widely used quality function. Values of $Q$ approaching 1 indicate partitions with strong community structure, while values close to 0 indicates that the partition does not correspond to a community structure.
Modularity has been extensively used both to evaluate and define community discovery algorithms for static and dynamic networks. 
However, the legitimacy of modularity has been challenged in recent years. 
In \cite{fortunato2007resolution}, the authors prove that partitions of optimal modularity do not necessarily correspond to what one expect as good communities, in particular, they introduced the problem of ``resolution limit," that may prevent from detecting small communities in large networks, and vice-versa. 

Indeed, modularity is not the only measure used to evaluate partition quality. 
Among the most used we can recall:
\begin{itemize}
    \item Conductance \cite{radicchi2004defining,Yang2012conductance}, i.e., the percentage of edges that cross the cluster border;
    \item Expansion \cite{radicchi2004defining}, i.e., the number of edges that cross the community border;
    \item Internal Density \cite{radicchi2004defining}, i.e., the ratio of edges within the cluster w.r.t. all possible edges;
    \item Cut Ratio and Normalized Cut \cite{fortunato2010community,shi2000normalized}, i.e., the fraction of all possible edges leaving the cluster;
    \item Maximum/Average ODF \cite{flake2000efficient} (out-degree fraction), i.e. the maximum/average fraction of nodes' edges crossing the cluster border;
    \item Flake ODF \cite{flake2000efficient}, i.e., the fraction of nodes involved in fewer edges within the community than outside it;
    \item Volume, i.e., the sum of degrees of nodes in the community;
    \item Edge Cut, i.e., the number of edges that should be removed to destroy all the paths connecting the community to the rest of the network.
\end{itemize}
All these indicators are used to get insights on the compactness and topological consistency of communities. 
Studies about the relations between these quality functions can be found in \cite{Yang2012conductance,creusefond2016evaluation}.
However, evaluating solutions based on a golden quality function has a major drawback: it favors methods that are designed to maximize it. 
Even though such strategy can be used fruitfully to compare methods that explicitly optimize a specific measure, its application to approaches that search for communities having different definition may produce misleading, or inconclusive/irrelevant, comparisons.\\

\noindent{\bf Internal.}
Several works overcome the issue of identifying a qualitative ranking among different algorithms by proposing a quantitative evaluation: instead of comparing with a ground-truth or computing quality measures, they focus on the algorithmic complexity \cite{Tantipathananandh2007,Bourqui2009,RPPG15}, running time \cite{LiLi2011,RPPG15,Gupta2012,SHY14}, scalability \cite{Folino2014,Tan2014,Zakrzewska2015} or analysis enabled by their approaches (i.e., identification of specific life-cycle events \cite{Palla2007,cazabet2012,Lee2014}). 

\emph{Internal} evaluations, conversely from external ones, do not want to produce a direct comparison among different partitions: they assume that each algorithm is based on a different community definition and -- after having described the structures each approach generates -- they measure quantitative performances and define a context of applicability for the proposed algorithm. 
Indeed, internal evaluation techniques are often used as support for external one, as in \cite{Agarwal2012,Folino2014,RPPG15}. 
The need for an evaluation strategy alternative to the external ones is primarily due to the intrinsic ill-posedness of the treated problem. 

As we discussed, community discovery in dynamic networks can be declined in several ways, by imposing different constraints on the network representation and its dynamics as well as focusing the analytical goal on different objectives.
This heterogeneous and fascinating scenario makes the evaluation task complex: unfortunately, so far complete and accepted methodologies able to address it were not presented, thus leaving researchers without a golden standard to follow.

\section{Applications}
\label{subsec:realdata}

Once categorized existing DCD approaches as well as the strategies adopted to validate their results, in this Section, we briefly discuss some related themes. 
In Section \ref{subsec:data}, the main categories of real-world dynamic network datasets are described, while in Section \ref{sec:visualization} visualization strategies for dynamic networks are introduced. 
Finally, in Section \ref{sec:DCDtool} examples of how DCD can be used as an analytical tool are provided.

\subsection{Real World Dynamic Networks }
\label{subsec:data}
The proliferation of online services, as well as the continuous collection of data regarding human activities, have given birth in recent years to the so-called \emph{Big Data} science. 
Indeed, almost all kinds of human-generated content can be used to model graphs: online social networks, mobility traces, information regarding collaborations or trade exchanges, web traffic are all examples of contexts in which it is possible to define relations among objects and design networks to analyze them. 
Unfortunately, so far only a few easily accessible online resources were made available to researchers to test dynamic network analytical tools. 
This situation is mostly due to the novelty of this research field: while static network datasets of any size (often enriched by valuable semantic annotations) have always been objects of analysis -- and thus collected and cataloged -- dynamic ones are often small, not well organized and rare.
Nevertheless, DCD approaches are often tested on such kind of datasets to evaluate their performances and compare runtimes on real-world scenarios.
In the following we discuss the main online and offline sources used to extract dynamic networks and provide a catalog of currently available datasets. 
\\ \ \\
\noindent{\bf Collaboration Networks.}
An extensively studied type of dynamic networks is the one that models work related collaborations, such as co-authoring of scientific publications or movie creation. 
Collaboration networks are extensively used by methods that work on snapshot graphs due to the natural timescale they provide (usually scientific papers -- as well as movies -- can be referred by their publication year). 
One drawback of networks built upon such kind of data is their lack of adherence to classical canons of real social networks: they are composed of interactions between groups more than activities between pairs of entities. 
Such characteristic reflects into networks structures composed of clique-chains whose density is usually higher than what is expected in traditional social networks. 
In Table \ref{tab:collab_net} are reported the most used online resources.\\ \

\begin{table}[!t]
\centering
\scriptsize
    \begin{tabular}{|l|l|m{5.4cm}|}        
        \multicolumn{3}{c}{\ \ }\\
        \hline
        {\bf Dataset} & {\bf Url} & {\bf Papers}\\
        \hline
        DBLP &  \url{http://goo.gl/5R9ozh} & \cite{Tang2008,Wang2008,ME09,Takaffoli2011,Goldberg2011,Huang2013}\\
        cond-mat & \url{http://arxiv.org} & \cite{Wang2008}\\
        cit-HepTh/Ph  & \url{http://arxiv.org} & \cite{Shang2012,Gorke2013,NS15}\\ 
        \hline
        IMDB & \url{http://imdb.com} & \cite{Goldberg2011}\\
        \hline
    \end{tabular}

 \caption{Dynamic Collaboration Networks}
 \label{tab:collab_net}
\end{table}

\begin{table}[b]
\centering
{
\scriptsize
    \begin{tabular}{|l|l|m{5.4cm}|}        
        \hline
        {\bf Dataset}  & {\bf Url} & {\bf Papers}\\
        \hline
        Twitter &  \url{http://twitter.com} & \cite{Lee2014}\\ 
        \hline
        Delicious.com &  \url{http://delicious.com} & \cite{Sun2010}\\
        \hline
    \end{tabular}
}
    \caption{Online/Offline Social Networks}
    \label{tab:soc_net}
\end{table}

\noindent{\bf Online/Offline Social Networks.}
The second class of sources often used to generate dynamic graphs is provided by online/offline social contexts. 
This kind of data is characterized by a clear semantic link: self-declared social relationships (that can be considered systematically reciprocated -- thus giving birth to undirected graphs -- or not necessarily so -- thus giving birth to directed graphs).
Social structures are often analyzed in different snapshots, whose durations are defined by the analyst since there is no natural temporal resolution to apply (as we have already discussed in Section \ref{sec:dyn_network} and in Section \ref{subsec:memory}). 
They can also be studied as temporal networks of relations.
In Table \ref{tab:soc_net} are listed some online datasources used to test algorithms discussed in Section \ref{sec:classification}. \\


\noindent{\bf Communication Networks.}
Data regarding communications (i.e., phone call logs, chats/emails, face-to-face interactions\dots) are often released with very precise timestamp annotations. 
Such fine-grained temporal information makes possible to build upon them not only snapshot graphs but also temporal networks of interactions, or of relations using a chosen TTL (Time To Live). 
Those datasets, listed in Table \ref{tab:com_net}, are primarily used by \textit{Partition update by set of rules} approaches (see Section \ref{sec:inf_iter}).\\

\begin{table}[!t]\centering
{
\scriptsize
    \begin{tabular}{|l |l |m{5.4cm}|}        
        \hline
        {\bf Dataset}  & {\bf Url} & {\bf Papers}\\
        \hline
        ENRON & \url{https://goo.gl/gQqo8l} & \cite{Falkowski2008,Tang2008,Wang2008,Folino2010,Folino2014,LiLi2011,Shang2012,Ferry2012} \\
        KIT & \url{https://goo.gl/gQqo8l} & \cite{Gorke2013} \\
        \hline
        Mobile phone & - & \cite{Folino2010,Folino2014,GJMZ12,Guo2014} \\
        \hline 
        wiki-Vote & \url{https://goo.gl/pzhst2} & \cite{Shang2012} \\
        Senate/Court Voting & - & \cite{CD15} \\
        \hline
        Thiers & \url{http://goo.gl/jhAcY8} & \cite{VLM15} \\
        CAIDA & \url{http://goo.gl/72y9iT} & \cite{ME09}\\
        \hline
        Facebook & \url{http://goo.gl/8J18cw} &\cite{RPPG15} \\
        Weibo & \url{http://goo.gl/bcoQiy} & \cite{RPPG15} \\
        Digg &  \url{http://goo.gl/ybKQS2} & \cite{Zakrzewska2015}\\
        Slashdot & \url{http://goo.gl/FGgEiM} & \cite{Zakrzewska2015}\\ 
        \hline
    \end{tabular}
    }
    \caption{Examples of real world dynamic graphs: Communication Networks }
    \label{tab:com_net}
\end{table}

\noindent{\bf Technological Networks.}
The last class of networks widely adopted to test dynamic algorithms is the technological one. 
Peer-to-Peer networks, as well as graphs built on top of Autonomous Systems routing information, are classical datasets, often used to validate even static community discovery approaches. 
In Table \ref{tab:tec_net} are shown some examples of technological networks used in revised algorithms. 

\begin{table}[!b]\centering
{

\scriptsize
    \begin{tabular}{|l|l|m{3.4cm}|}        
        \hline
        {\bf Dataset} & {\bf Url} & {\bf Papers}\\
        \hline
        AS Routers & \url{https://goo.gl/tGk1lJ} & \cite{Gorke2013,AHS14} \\
        p2p-Gnutella & \url{https://goo.gl/Tgl5NC} & \cite{Huang2013}\\
        \hline
    \end{tabular}
    }
    \caption{Technological Networks}
    \label{tab:tec_net}
\end{table}

\subsection{Visualization}
\label{sec:visualization}
Dynamic network visualization is a challenging task: so far, to the best of our knowledge, few solutions have been proposed to visualize the evolution of communities. 
Effectively represent community dynamics is difficult since it needs to combine the challenges presented by both evolving network and communities visualization. For a survey on the visualization of dynamic networks, please refer to \cite{CGF:CGF12791}. For a survey on the visualization of community structures, please refer to \cite{vehlow2015state}.

We can classify visualization methods in two categories: (i) dynamic visualizations, such as videos, in which the state of communities at different time steps is represented by independent visualizations and (ii) static drawings that summarize the whole network/communities evolution.

\subsubsection{Dynamic Visualization}
Methods falling in this category have been designed to represent communities in static networks, and have been extended to handle dynamic networks by the means of animations. 
This adaptation usually consists in using a layout of nodes adapted to favor the stability of their position despite the changes in their connectivity.

\cite{frishman2004dynamic} focus on minimizing the change of positions of each cluster as a whole, and of the nodes within them. 
Invisible nodes representing communities are used to ensure the stability of community centroids. 
Communities are represented by node coloring, and/or by drawing a square around nodes belonging to the same group.

In \cite{beiro2010visualizing}, the whole community is represented as a single node, whose size corresponds to its number of components. 
Edges between communities represent all the links that connect their respective nodes. 
In \cite{boyack2009mapping}, communities are shown as pie-charts, a visualization used to summarize some property of the nodes they are composed of.

\begin{figure}[!t] 
\centering
\includegraphics[width=0.4\textwidth]{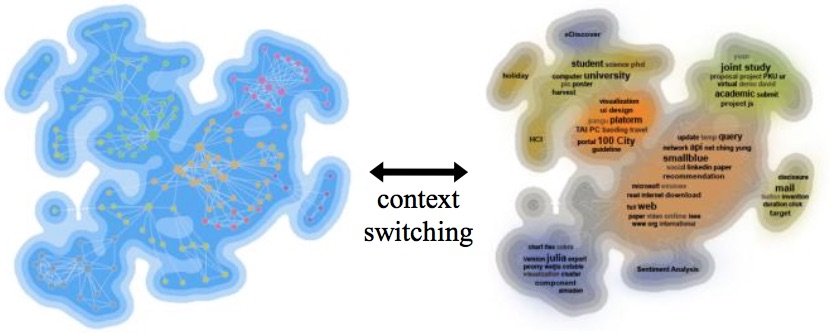} 
\caption{Visusalisation using contour map on top of nodes positioned with an appropriate layout. Image taken from: \cite{lin2010contextour} (Copyright ©2010 Society for Industrial and Applied Mathematics.  Reprinted with permission.  All rights reserved.)
 } \label{img:contourMap} 
\end{figure}

\cite{hu2012embedding} proposes to position nodes in the network based on a classic node positioning layout, and then overlay a shape on top of the nodes belonging to the same community, in a manner appropriate for an easy understanding of the community structure, inspired by geographical maps. 
In \cite{lin2010contextour} a similar approach is used, but a fuzzy contour-map is used to represent communities. (Fig. \ref{img:contourMap})

\subsubsection{Static Visualization}
\begin{figure}[!b]
\centering
\includegraphics[width=0.4\textwidth]{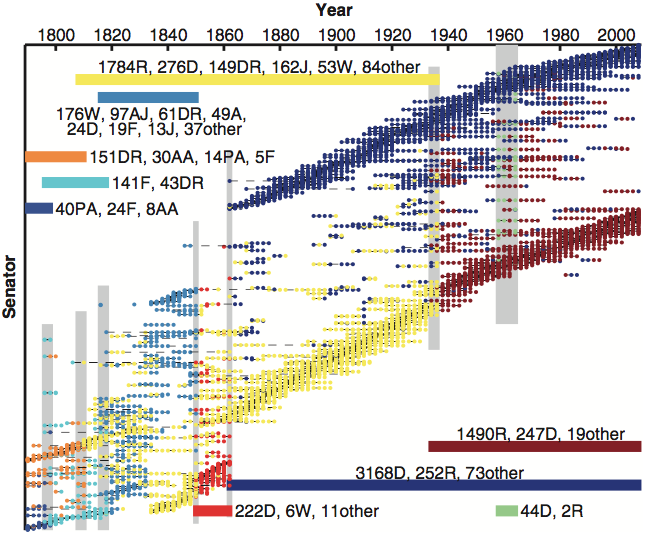} 
\caption{Visusalisation used by Mucha et al. Image taken from: \cite{Mucha2010}}
\label{img:muchaEtAl} 
\end{figure}

In \cite{Mucha2010}, a vertical position is attributed to each node, according to its first apparition and its community affiliation, in order to put similar nodes close to each other. 
The horizontal axis corresponds to time. 
A color is attributed to each community, and, for each timestamp, the node, if present, is represented by a dot colored according to the community it belongs to (see Fig. \ref{img:muchaEtAl}).

In \cite{reda2011visualizing}, each community is represented as a fixed-width horizontal space and nodes as colored lines that switch between them when community affiliation changes occur.

A similar idea is proposed in \cite{rosvall2010mapping}. 
There, alluvial diagrams are used to represent community evolution: each node is represented as a continuous line, whose vertical position is determined by its community affiliation. 
Nodes belonging to the same community at a given timestamp always appear next to each other and sharing the same color. 
Therefore, communities appear as horizontal strips whose width corresponds to the number of nodes they contain.

\begin{figure}[t] 
\centering
\includegraphics[width=0.7\textwidth]{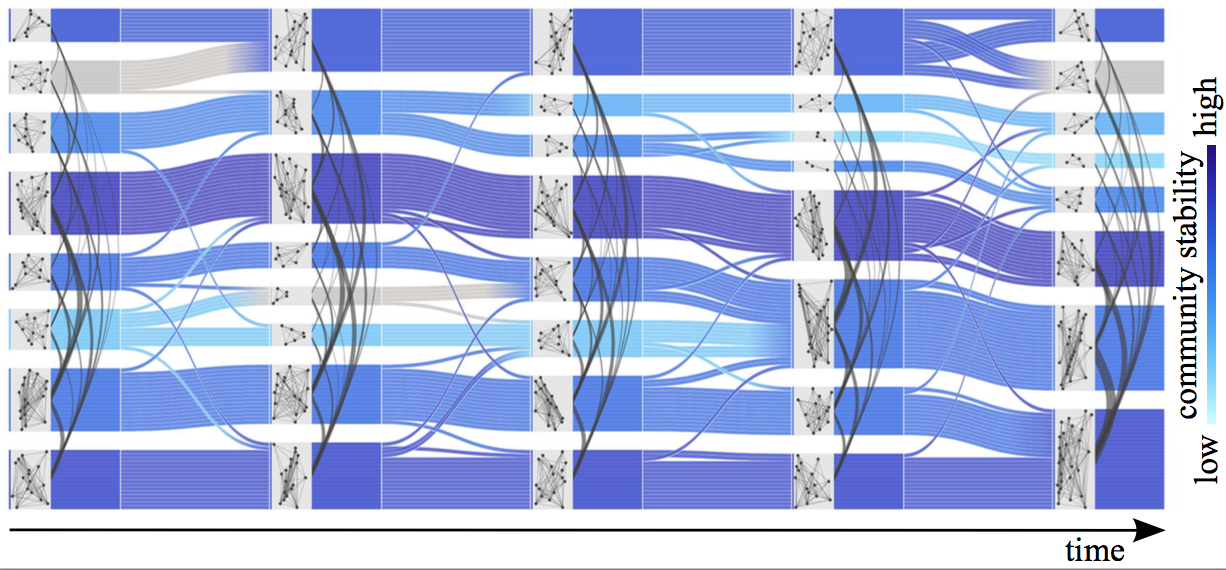} 
\caption{Visusalisation by alluvial diagrams. Image taken from: \cite{vehlow2015visualizing} }
\label{img:alluvial} 
\end{figure}

The authors of \cite{vehlow2015visualizing} propose to combine views of the structure of the communities with the alluvial representation. 
They employ a custom sorting of the communities and vertices per time step to minimize crossing, and an automatic color assignment methods to improve readability (see Fig. \ref{img:alluvial}).

\begin{figure}[t] 
\centering
\includegraphics[width=0.7\textwidth]{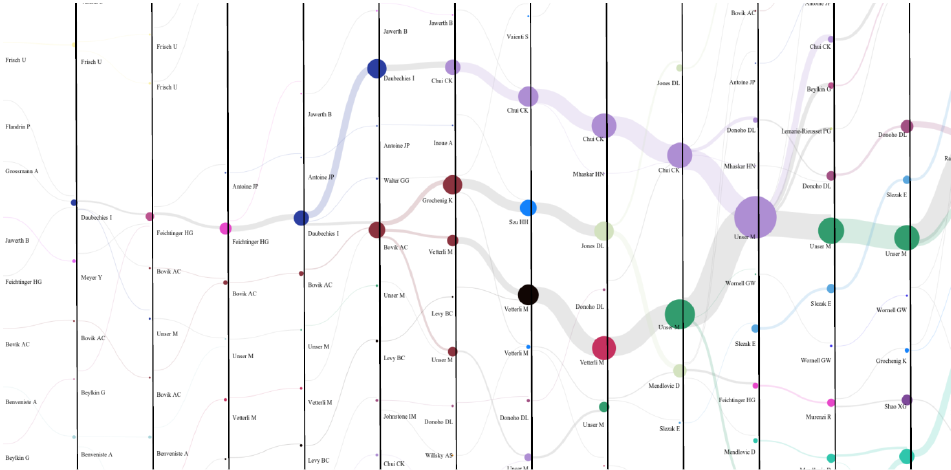} 
\caption{Variation of the alluvial diagram to improve the readability of the life-cycle of communities. Image taken from: \cite{morini2015temporal} }
\label{img:alluvialMorini} 
\end{figure}

The authors of \cite{morini2015temporal} use a variation of the alluvial diagram in which the vertical position of communities is less constrained, facilitating the tracking of communities life-cycles.( see Fig. \ref{img:alluvialMorini})

In \cite{Vehlow2016}, a method is proposed to visualize both a hierarchical organization of communities and their evolution through time. The method use matrices to represent each step, and the relations between communities in different time steps or different hierarchical levels are represented by edges between the communities.

\label{sec:applications}
\subsection{DCD as a tool}
\label{sec:DCDtool}
Several articles \cite{Agarwal2012,Lee2014,cazabet2012} proposed to use DCD algorithms to identify events in online social networks. 
The general approach they follow consist of creating a dynamic network of keywords or n-grams, in which edges correspond to a semantic relationship between terms observed in the studied OSN. 
The dynamic communities found on such networks correspond to ``popular events" aggregating similar bursting terms.
The evolution of groups of people in social media has also been addressed in \cite{gliwa2012identification,Saganowski2015}, on a Polish blogosphere service, Facebook, and DBLP.

DCD approaches could also be used as a tool for telecommunication networks. 
As such methods allow to keep up-to-date information on the community structure in real-time, they enable for the adoption of algorithms able to extract knowledge from highly dynamical network substructures. 
In \cite{Nguyen2011b}, the authors propose to use DCD for improving routing in MANETs (Mobile Ad Hoc Networks), while in \cite{Nguyen2011}, the authors propose to use them to contain worms, notably by concentrating on the immunization of nodes at the interface between the infected community and the rest of the network. 
In \cite{schlitter2009mining}, the authors study the music listened by a set of 1800 users on the platform \emph{Last.fm} over a period of 3 years.
They detect groups of users capturing specific music genres and  study their evolution. 
For instance, they observe the fusion between a group of users listening to progressive metal and another group listening to death metal.

Scientific collaboration network analysis is a topic on which DCD algorithms have been applied several times, notably in \cite{Cazabet2010,Lin2009,Nguyen2012,rosvall2010mapping}. 
One of the most in-depth analysis can be found in \cite{rosvall2010mapping}, where a merge between the fields of Neurology, Psychology and some areas of Molecular and Cell biology is identified, forming a new field of Neuroscience between 2003 and 2005.

In \cite{Braun15092015}, DCD is used to study the reconfiguration of the brain network during executive cognition in humans.

Finally, in \cite{Mucha2010}, political affiliation among senators is studied based on vote similarities. 
Known historical changes in U.S politics are compared with communities of senators found by a dynamic community discovery algorithm. The same method has been applied for the United Nations General Assembly \cite{MACON2012343}, exploring different processes to construct the network from data.

\section{Outlook}
\label{sec:conclusion}
Despite the significant amount of literature already published on the topic of Dynamic Community Discovery, numerous lines of research  remain open. 
The problem \emph{per se} is not trivial: indeed, we have seen that it cannot be reduced to simply running a static algorithm at several points in time. 

As the scientific community is still tackling the problem of what is a good community, researchers on DCD are confronted with the problem of defining what it means for a community to persist and evolve. 
Moreover, they need to solve the apparent paradox that ``optimal" partitions found during different network evolution steps do not always form a coherent dynamic structure.

An underlying problem is due to the heterogeneous nature of the dynamic networks used by different methods. 
While snapshots and temporal networks have the same expressivity -- i.e., both of them can be used to model the same network topology dynamics -- they often have to be considered differently in practice. 
Snapshots are often used when the network varies greatly between subsequent observations, while temporal networks represent a fine-grained evolution. 
As a consequence, a method designed for one of these approaches often solve a different problem than one designed for the other.

The classification we have adopted highlights the amazing variety of approaches adopted by different authors. 
Such DCD methodologies do not only propose different ways of solving the same problem, but also explore different definitions of what is a dynamic community, and how to find network partitions which are both coherent on the long run and meaningful at given points in time.

Despite this variety of definitions and approaches, common grounds have begun to emerge, in particular, the definition of \textit{events} or \textit{operations}.

\bigbreak

What we think the field lacks the most is a common ground to evaluate and compare the different methods among themselves. 
Indeed, earliest methods for Community Detection have compared themselves to the Zachary Karate Club, while there is no such widespread toy model for dynamic networks.

More importantly, the field of CD has been structured by the introduction of advanced benchmark and systematic, quantitative comparison of algorithms. 
While a few steps have been done in this direction, by introducing simplistic synthetic benchmarks, there is still no universally recognized benchmark equivalent to the LFR for the dynamic case, and no systematic comparison of methods has been conducted yet. 
The problem is indeed arduous: to compare methods and their results, it is mandatory to provide to each method an appropriate input, and convert outputs to comparable pieces of information. 
Indeed, the problem is not only a technical one: it involves the lack of standard formal definitions. 
How to compare methods working on a few snapshots and methods on temporal networks? How to compare results that do not share the same possible operations of communities? And how to design a benchmark that generates realistic evolution of communities while such evolution is not well known yet?

A question many will ask is: what method should I use for my data? 
Unfortunately, there is not a final answer to this question, in particular, due to the lack of systematic evaluation.
Nevertheless, the classification of methods according to their characteristics already give us some hints. 
A first aspect to consider is the number of steps of evolution present in the original data. 
If there are few steps, (i.e. less than 10), typically corresponding to aggregated observations over months or years, one should in priority consider approaches based on snapshots, while if the number of steps is high (1000s or more), methods using temporal networks are more suitable. 
Another aspect is the priority given to the quality of communities found at a particular instant rather than to the coherence of the continuity of communities. 
For instance, a tool using DCD to improve in real-time the routing in a point to point network might give a higher importance to the current state of the network. 
Conversely, a retrospective analysis of past events aimed at interpreting the emergence of scientific fields by clustering journals or papers should favors the long-term coherence of communities.

Finally, we observed that despite a large number of methods proposed,  few applications of DCD to concrete problems were proposed -- except the ones by the authors themselves. 
As of today, it seems that the field lacks a visibility and that authors who need a DCD method tend to develop their own, ad-hoc algorithm, often without considering existing ones. 
We think that the future of the field will rely on its greater cohesion and ease of access for newcomers, to reach an audience as large as conventional Community Discovery.

\section*{Acknowledgement}
We thank Jane Carlen for her feedback that helped us correct some errors in the description of some methods.
\bibliographystyle{apalike}
\bibliography{biblio}


\clearpage

In this section, we propose a brief description for each DCD method we have identified in the literature, framing them in the proposed taxonomy. 
When possible, we contacted the authors to validate our description. 
Inside categories, methods are ordered chronologically.
\section{Instant-optimal Community Discovery}
Here are collected and classified all the works introducing approaches falling in the \emph{Instant-optimal} class.
For details on the class definition and peculiarities refer to Section \ref{subsec:instantoptimal}.

\subsection{Iterative similarity-based approaches} 
\label{subcat1}

The algorithm in \cite{Hopcroft2004} can be seen as the ancestor of all approaches within this category. 
It first finds ``natural communities" in each snapshot by running the same algorithm for community detection several times and maintaining only those communities that appear in most runs. 
In a second phase, similar communities of successive snapshots are matched together using a match function defined as:
\[
match(C,C') = Min\left(\frac{|C \cap C'|}{|C|},\frac{|C\cap C'}{|C'|}\right)
\]
with $C$ and $C'$ the clusters to compare treated as sets.
\\ \ \\
The algorithm proposed in \cite{Palla2007} is based on clique percolation (CPM \cite{palla2005uncovering}). 
CPM is applied on each snapshot to find communities. 
The matching between these communities is done by running CPM on the joint graph of snapshots at $t$ and $t+1$. 
Due to the formal definition of communities provided by CPM (communities are composed of cliques), communities of the joint graph can only grow compared to those of both snapshots.
The authors also discuss how the knowledge of the temporal commitment of members to a given community can be used for estimating the community's lifetime. 
They observed that while the stability of small communities is ensured by the high homogeneity of the nodes involved, for large ones the key of a long life is continuous change.
\\ \ \\
In \cite{Bourqui2009}, an approach based on dynamic graph discretization and graph clustering is proposed. 
The framework allows detection of major structural changes over time, identifies events by analyzing temporal dimensions and reveals hierarchies in social networks. 
It consists of four major steps:
\begin{enumerate}
    \item The social network is decomposed into a set of snapshot graphs;
    \item Clusters are extracted from each static graph separately using an overlapping CD algorithm, to produce overlapping structures. 
    This step allows to identify communities in the network but also its pivots (vertices shared by several clusters) while being insensitive to minor changes in the network;
    \item Major structural changes in the network are detected by comparing the clustering obtained on every pair of successive static graphs using a similarity measure (\emph{cluster representativeness}, e.g., the geometrical mean of the normalized ratio of common elements between two clusters). 
    Thus, the temporal changes in the input network are decomposed into periods of high activity and \emph{consensus} communities during stable periods;
    \item An influence hierarchy in the \emph{consensus} communities is identified using an MST (Minimum Spanning Tree). 
\end{enumerate}
In \cite{Asur2007}, the authors approach the problem of identifying events in communities: MCL (a modularity based algorithm) is applied to extract communities in each time step.  
A set of events (Continue, kMerge, kSplit, Form, Dissolve, Join, Leave) are defined. For instance, kMerge is defined as: $kMerge(C^k_i,C^l_i,k)=1$ iff $\exists C^j_{i+1}$ such that
\[
 \frac{|(V^k_i \cup V^l_i) \cap V^j_{i+1}|}{Max(|V_i^k \cup V^l_i|,|V^j_{i+1}|)}>k\%
\]
and $|V^k_i \cap V^j_{i+1}|>\frac{|C^k_i|}{2}$ and $|V^l_i \cap V^j_{i+1}|>\frac{|C^l_i|}{2}$,
with $i$ representing time steps.
\\ \ \\ 

In \cite{rosvall2010mapping}, the authors propose a method to map changes in large networks. 
To do so, they focus on finding stable communities for each snapshot, by using a bootstrap resampling procedure accompanied by significance clustering. 
More precisely, for each snapshot, 1000 variations of the original network are generated, in which the weight of each edge is taken randomly from a Poisson distribution having as mean the original weight of the edge. 
All of these alternative networks are clustered using a static algorithm. 
To find significant communities, the authors propose to search for modules, largest sets of nodes that are clustered together in more than 95\% of networks. 
These small node sets are grouped into larger clusters if their nodes appear together in more than 5\% of network instances. 
Clusters corresponding to different time steps are linked if the number of common nodes is higher than a user-defined threshold. 
\\ \ \\
In \cite{BKP11}, the authors propose an extension of the basic community events described in \cite{Palla2007} and an approach capable of handling communities identified by a non-monotonic community detection algorithm ($N^{++}$ \cite{bota2010community}). 
The proposed method works on snapshots and uses for each pair of adjacent network observation \emph{union} (and \emph{intersection}) graphs to perform community matching. 
The final goal of this approach is to build an extended community life-cycle. 
In addition to the classical events defined in \cite{Palla2007}, 5 meta-events are introduced to capture the combinations of standard events: Grow-merge, Contraction-merge, Grow-split, Contraction-split and Obscure case (which envelop more complex, non-binary, transformations).
\\ \ \\
The method proposed in \cite{Takaffoli2011} is designed to track the evolution of communities. 
First, a static community detection is run on each snapshot, using an existing algorithm. 
In a second step, communities matching and operations are done according to a set of rules, that can be summarized as:
\begin{itemize}
    \item Two communities are matched if they have at least a fraction $k$ of the nodes of the largest one in common;
    \item A new community is born if it has no match in the previous snapshot;
    \item A community dies if it has no match in the following snapshot;
    \item A community split if several other communities have more than a fraction $k$ of its nodes;
    \item A merge occurs if a community contains at least a fraction $k$ of several other communities.
\end{itemize}

The method proposed in \cite{greene2010tracking}  is designed to track the evolution of communities in time evolving graphs following a two-step approach: (i) community detection is executed on each snapshot with a chosen static algorithm and then (ii) the Jaccard coefficient is used to match communities across adjacent snapshots. 
Communities are matched if their similarity is above a user-defined threshold, allowing many-to-many mappings.
\\ \ \\
The GED method, proposed in \cite{Brodka2013}, is designed to identify what happened to each node-group in successive time frames: it uses not only the size and equivalence of groups' members but also takes into account their position (e.g., via node ranking within communities, social position\dots) and importance within the group. 
This approach is sustained by a new measure called \emph{inclusion}, which respects both the \emph{quantity} (the number of members) and \emph{quality} (the importance of members) of the group. 
The proposed framework is parametric in the static CD algorithm to be applied on each network snapshot to extract node-groups (communities).
\\ \ \\
The method proposed in \cite{Dhouioui2014} uses the OCDA algorithm \cite{Dhouioui2013} to find overlapping communities in each snapshot. 
Then, relations are found between similar communities in different snapshots. 
A Data Warehouse layer memorizes the evolution of the network and its communities.
\\ \ \\
In \cite{NS15} an evolving network is represented by a series of static snapshots. 
Communities are determined using the Louvain algorithm \cite{blondel2008fast}, then matched using a custom similarity metric. 
The evolution of communities is labeled with the events they experienced (survive, grow, shrink, merge, split and dissolve). 
Moreover, the authors propose to forecast future evolution of communities. 
To do so, a broad range of structural and temporal features are extracted covering many properties of both the internal link structure and the external interaction of each community with the rest of the network.  
A temporal window length is defined to generate time series for each feature, thus for each community, a set of time-series is built. 
Then, ARIMA (Auto Regressive Integrated Moving Average) model \cite{newbold1983arima} is applied to estimate the next values of the features.
The communities with the forecasted feature values were then used as test set for several classification algorithms trained using the rest of the window length snapshots as the training set. 

\subsection{Iterative core-nodes based approaches}
\label{subcat2}

The method described in \cite{Wang2008} proposes to use the concept of core nodes to track the evolution of communities. 
It is agnostic w.r.t. the CD algorithm used to detect communities on each snapshot. 
For each snapshot, a voting strategy is used to select a few core nodes. 
Finally, the tracking of communities can be done by tracking the core nodes.
\\ \ \\
In \cite{Chen2010}, the authors adopt a conservative definition of communities, i.e., they define them as the maximal cliques of the graph. 
They consequently introduce \emph{Graph representatives} and \emph{Community representatives} to limit the search space and avoid redundant communities. 
They are defined as follows:
\begin{itemize}
    \item \emph{Graph representatives}: Representatives of graph $G_t$ are the nodes that also appear in $G_{tâ1}$, $G_{t+1}$,or both. 
    Nodes that only appear in one graph are called graph dependent nodes. 
    If a community only contains graph-dependent vertices, then it can be considered as a ``graph-dependent" community, not dynamic one
    \item \emph{Community representatives}: A community representative of community $C_i^t$ is a node in $C_i^t$ that has the minimum number of appearances in other communities of the same graph $G_t$.
\end{itemize}
The proposed approach first finds \emph{graph representatives} and enumerates the communities that are seeded by the graph representatives to avoid considering redundant communities. 
Then, in every generated community, it selects only one node as a community representative and uses community representatives to establish the predecessor/successor relationship between communities of different time-steps. 
Once all predecessors and successors have been found, it finally applies decision rules to determine community dynamics. 

\subsection{Multi-step matching}
\label{subcat3}

In \cite{Falkowski2006,FS07}, a three-step approach is applied to detect subgroups in social networks:
\begin{enumerate}
    \item In the first step, communities are found in each snapshot using a static CD algorithm.
    \item In the second step, communities in different snapshots are linked based on their similarity, using the overlap measure: $overlap(x,y)=\frac{|x\bigcap y|}{min(|x|,|y|)}$. This creates a \textit{Community survival graph}.
    \item In the third step, a community detection algorithm is run on the \textit{Community survival graph}, thus finding communities relevant across several snapshots.
\end{enumerate}

The approach proposed in \cite{Goldberg2011} identifies evolutive chains of communities.
Given a time-evolving graph, community detection on each snapshot is executed using a chosen static algorithm (including overlapping ones). 
Any set intersection based measure can be used to match communities between snapshots, creating a link between them. 
The authors propose a strategy to find the best chains of evolution for each community: they define chain strength as the strength of its weakest link. 
As a result, all the maximal valid chains are constructed for the identified communities. 
A valid chain is considered maximal if it is not a proper subchain of some other valid chain.
\\ \ \\
In \cite{morini2017revealing}, the authors start by computing aggregated network snapshots from timestamped observations (SN or TN) using sliding windows. Communities are detected using a static CD algorithm on each of these windows independently. Then, the similarity between pair of communities is computed between communities at $t$ and communities at $t-2$, $t-1$, $t+1$, $t+2$. This information is used to \textit{smooth out} the evolution of communities: if a community $c_{t-n}$ in a previous snapshot has a high similarity with $c_{t+n}$ in a later one, but is matched with lower similarity to two communities $c1_{t}$ and $c2_{t}$ at $t$, then these two communities are merged and the resulting community is identified to $c_{t-n}$ and $c_{t+n}$. The same mechanism is used to smooth out artificial merges.

\section{Temporal Trade-off Communities Discovery}
Here are collected and classified all the works introducing approaches falling in the \emph{Temporal Trade-off} class.
For details on the class definition and peculiarities refer to Section \ref{sec:inf_iter}.

\subsection{Partition update by Global optimization}
\label{subcat4}

In \cite{ME09}, the authors adapt a dynamic extension of Latent Dirichlet Allocation to the extraction of dynamic communities. 
They design {\scshape cDTM-G} the ``continuous time Dynamic Topic Model for Graphs".
In this work, observations of the time evolving graph (snapshots) correspond to \emph{corpora} at different points in time: each source-node at each time step corresponds to a document at the same time step, and the links among the nodes connect document's words. 
Exploiting such modeling strategy, dynamic groups/communities capture time evolving topics.
The authors sequentially run LDA-G \cite{henderson2009applying} on each time step initializing the topics for the current time step with the ones learned in the previous time step.
\\ \ \\
The method proposed in \cite{aynaud2010static} is based on the Louvain algorithm. 
In the original, static version, the Louvain algorithm is initialized with each node in its community. 
A greedy, multi-step heuristic is used to move nodes to optimize the partition's modularity. 
To limit instability, the authors propose to initialize the Louvain algorithm at $t$ by communities found at $t-1$. 
A parameter $x \in [0-1]$ can be tuned to specify a random fraction of nodes that will start in a singleton community, to avoid staying in a local minimum (community drift). 
\\ \ \\
The algorithm proposed in \cite{Gorke2010} efficiently maintains a modularity-based clustering of a graph for which dynamic changes arrive as a stream. 
The authors design dynamic generalizations of two modularity maximization algorithms, \cite{clauset2004finding} and \cite{blondel2008fast}. 
The problem is cast as an ILP (Integer Linear Programming). 
Intuitively, the process at $t$ is initialized by communities found at $t-1$, adding a backtracking strategy to extend the search space.
\\ \ \\
In \cite{Bansal2011}, the authors introduce a dynamic community detection algorithm for real-time online changes, which involves the addition or deletion of edges in the network. 
The algorithm, based on the greedy agglomerative technique of the CNM algorithm \cite{clauset2004finding}, follows a hierarchical clustering approach, where two communities are merged at each step to optimize the increase in the modularity of the network. 
The proposed improvement for the dynamic case consists in starting from the dendrogram of the previous step cutted just before the first merge of a node involved in the network modifications. 
It assumes that the small change in network size and density between snapshots does not dramatically impact modularity for non-impacted nodes, and therefore that the beginning of the dendrogram is not impacted.
\\ \ \\
The method proposed in \cite{Shang2012} searches for evolving communities of maximal modularity. 
The Louvain algorithm \cite{blondel2008fast} is used to find communities in the first snapshot. 
Custom rules are then applied to each modification of the network to update the communities while keeping the modularity as high as possible. 
The procedure designed to update the communities depends on the type of modification of the network, which can concern an inner community edge, across community edge, a half-new edge (if one of the involved nodes is new) or a new edge (if both extremities are new nodes).
\\ \ \\
The authors of \cite{AHS14} introduce a game-theoretic approach for community detection in dynamic social networks called D-GT: in this context, each node is treated as a rational agent who periodically chooses from a set of predefined actions to maximize its utility function. 
The community structure of a snapshot emerges after the game reaches Nash equilibrium: the partitions and agent information are then transferred to the next snapshot. 
D-GT attempts to simulate the decision-making process of the individuals creating communities, rather than focusing on statistical correlations between labels of neighboring nodes.

\subsection{Informed CD by Multi-objective optimization}
\label{subcat5}

In \cite{Zhou2007}, the authors seek communities in bipartite graphs. 
Temporal communities are discovered by threading the partitioning of graphs in different periods, using a constrained partitioning algorithm. 
Communities for a given snapshot are discovered by optimizing the \textit{Normalized cut}.
The discovery of community structure at time $t$ seeks to minimize the (potentially weighted) sum of distances between the current and $n$ previous community membership, $n$ being a parameter of the algorithm.
\\ \ \\ 
The approach proposed in \cite{Tang2008} allows finding the evolution of communities in multi-mode networks. 
To find communities at time $t$, the authors propose to use a spectral-based algorithm that minimize a cost function defined as the sum of two parts, $F_1$ and $\Omega$. 
$F_1$ corresponds to the reconstruction error of a model similar to block modeling but adapted to multi-mode networks. 
$\Omega$ is a regulation term that states the difference between the current clustering and the previous one. 
Weights allow balancing the relative importance of these two factors.
\\ \ \\
In \cite{Yang2009,Yang2010}, the authors propose to use a Dynamic Stochastic Block Model. 
As in a typical static SBM, nodes belong to clusters, and an interaction probability $\beta_{ql}$ is assigned to each pair of clusters $(q,l)$. 
The dynamic aspect is handled by a transition matrix, that determines the probability for nodes belonging to a cluster to move to each other cluster at each step. 
The optimal parameters of this model are searched for using a custom Expectation-Maximization (EM) algorithm. 
The $\beta_{ql}$ are constant for the same community for all time steps.

Two versions of the framework are discussed:
\begin{itemize}
    \item \emph{online learning} which updates the probabilistic model iteratively (in this case, the method is a Temporal Trade-Off CD, similar to FacetNet \cite{Lin2009});
    \item \emph{offline learning} which learns the probabilistic model with network data obtained at all time steps (in this case, the method is a Cross-Time CD, see appropriate section).
\end{itemize}
The method discussed in \cite{Folino2010,Folino2014} adopt a genetic algorithm to optimize a multi-objective quality function. 
One objective is to maximize the quality of communities in the current snapshot (the modularity is used in the article, but other metrics such as conductance are proposed). 
The other objective regards the maximization of the NMI between the communities of the previous step and of the current one to ensure a smooth evolution.
\\ \ \\
In FacetNet, introduced in \cite{Lin2009,Lin2008}, the authors propose to find communities at $t$. 
To minimize the reconstruction error, a cost error defined as $cost = \alpha C_1 + (1-\alpha) C_2$ is defined -- with $C_1$ corresponding to the reconstruction error of a mixture model at $t$, and $C_2$ corresponding to the KL-divergence between clustering at $t$ and clustering at $t-1$. 
The authors then propose a custom algorithm to optimize both parts of the cost function. 
This method is similar to \cite{Yang2009}.
\\ \ \\
The method described in \cite{Sun2010,sun2014co} proposes to find multi-typed communities in dynamic heterogeneous networks. 
A Dirichlet Process Mixture Model-based generative model is used to model the community generations. 
A Gibbs sampling-based inference algorithm is provided to infer the model. 
At each time step, the communities found are a trade-off between the best solution at $t$ and a smooth evolution compared to the solution at $t-1$.
\\ \ \\
In \cite{GJMZ12}, the detection of community structure with temporal smoothness is formulated as a multi-objective optimization problem. 
As in \cite{Folino2010,Folino2014} the maximization is performed on modularity and NMI.
\\ \ \\
In \cite{Kawadia2012} a new measure of partition distance is introduced.
\emph{Estrangement} captures the inertia of inter-node relationships which, when incorporated into the measurement of partition quality, facilitates the detection of temporal communities.
Upon such measure is built the \emph{estrangement confinement method}, which postulates that neighboring nodes in a community prefer to continue to share community affiliation as the network evolves.
The authors show that temporal communities can be found by estrangement constrained modularity maximization, a problem they solve using Lagrange duality. 
Estrangement can be decomposed into local single node terms, thus enabling an efficient solution of the Lagrange dual problem through agglomerative greedy search methods.
\\ \ \\
In \cite{CD15} is introduced a hidden Markov model for inferring community structures that vary over time according to the cut-and-paste dynamics from \cite{C14}. 
The model takes advantage of temporal smoothness to reduce short-term irregularities in community extraction. 
The partition obtained is without overlap, and the network population is fixed in advance (no node appearance/vanishing).
\\ \ \\
In \cite{Gorke2013} is proposed an algorithm to efficiently maintain a modularity-based clustering of a graph for which dynamic changes arrive as a stream, similar to \cite{Gorke2012}.
Differently from such work, here the authors introduce an explicit trade-off between modularity maximization and similarity to the previous solution, measured by the Rand index. 
As in other methods, a parameter $\alpha$ allows to tune the importance of both components of this quality function.

\subsection{Partition update by set of rules}
\label{subcat6}

DENGRAPH \cite{Falkowski2008} is an adaptation of the data clustering technique DBSCAN \cite{ester1996density} for graphs.
A function of proximity ranging in [0,1] is defined to represent the distance between each pair of vertices. 
A vertex $v$ is said to be density-reachable from a core vertex $c$ if and only if the proximity between $v$ and $c$ is above a parameter $\omega$.
A core vertex is defined as a vertex that has more than $\eta$ density-reachable vertices.
Communities are defined by the union of core vertices that are density-reachable from each other.
A simple procedure is defined to update the communities following each addition or removal of an edge.
\\ \ \\
AFOCS (Adaptative FOCS), described in \cite{Nguyen2011}, uses the FOCS algorithm to find initial communities. 
Communities found by FOCS are dense subgraphs, that can have strong overlaps. 
Starting from such topologies, a local procedure is applied to update the communities at each modification of the network (e.g., addition or removal of a node or an edge). 
The strategy, proposed to cope with local network modifications, respects the properties of communities as they are defined by the FOCS algorithm.
\\ \ \\
In \cite{Cazabet2010}, is introduced iLCD, an incremental approach able to identify and track dynamic communities of high cohesion. 
Two measures are defined to qualify the cohesion of communities:
\begin{itemize}
    \item EMSN: the Estimation of the mean number of second neighbors in the community;
    \item EMRSN: the Estimation of the mean number of robust second neighbors (second neighbors that can be accessed by at least 2 different paths).
\end{itemize}
Given a new edge, the affected nodes can join an existing community if an edge appeared with a node in this community. 
The condition is that the number of its second neighbors in the community is greater than the value of EMSN of the community, and respectively for EMRSN.
iLCD has two parameters $k$ and $t$. 
The former regulates the rising of new communities (new communities are created if a new clique of size $k$ appears outside of existing ones), while the latter the merge of existing ones (communities can be merged if their similarity becomes greater than a threshold $t$).
\\ \ \\
In \cite{Nguyen2011b}, the Quick Community Adaptation (QCA) algorithm is introduced. 
QCA is a modularity-based method for identifying and tracking community structure of dynamic online social networks. 
The described approach efficiently updates network communities, through a series of changes, by only using the structures identified from previous network snapshots. 
Moreover, it traces the evolution of community structure over time. 
QCA first requires an initial community structure, which acts as a \emph{basic structure}, to process network updates (node/edge addition/removal).
\\ \ \\
In \cite{cazabet2011simulate}, the authors propose to use a multi-agent system to handle the evolution of communities. 
The network is considered as an environment, and communities are agents in this environment. 
When what they perceive of their local environment change (edge/node addition/removal), community agents take decisions to add/lose nodes or merge with another community, based on a set of rules. 
These rules are based on three introduced measures, \textit{representativeness}, \textit{seclusion} and \textit{potential belonging}.
\\ \ \\
In \cite{Agarwal2012} the authors addressed the problem of discovering events in a microblog stream. 
To this extent, they mapped the problem of finding events to that of finding clusters in a graph. 
The authors describe aMQCs: bi-connected clusters, satisfying \emph{short-cycle property} that allow them to find and maintain the clusters locally without affecting their quality. 
Network dynamics are handled through a rule-based approach for addition/deletion of nodes/edges. 
\\ \ \\ 
In \cite{Duan2012}  social networks' dynamics are modeled as a \emph{change stream}. 
Based on this model, a local DFS forest updating algorithm is proposed for incremental 2-clique clustering, and it is generalized to incremental k-clique clustering. 
The incremental strategies are designed to guarantee the accuracy of the clustering result with respect to any kind of changes (i.e., edge addition/deletion). 
Moreover, the authors shown how the local DFS forest updating algorithm produces not only the updated connected components of a graph but also the updated DFS forest, which can be applied to other issues, such as finding a simple loop through a node.
\\ \ \\
In \cite{Gorke2012}, the authors show that the structure of \emph{minimum-s-t-cuts} in a graph allow for an efficient dynamic update of \emph{minimum-cut trees}.
The authors proposed an algorithm that efficiently updates specific parts of such a tree and dynamically maintains a graph clustering based on minimum-cut trees under arbitrary atomic changes. 
The main feature of the graph clustering computed by this method is that it is guaranteed to yield a certain \emph{expansion} -- a bottleneck measure -- within and between clusters, tunable by an input parameter $\alpha$. 
The algorithm ensures that its community updates handle temporal smoothness, i.e., changes to the clusterings are kept at a minimum, whenever possible.
\\ \ \\
The CUT algorithm, introduced in \cite{Huang2013}, tracks community-seeds to update community structures instead of recalculating them from scratch as time goes by. 
The process is decomposed into two steps:
\begin{enumerate}
    \item First snapshot: identify community seeds (collection of 3-cliques); 
    \item Successive snapshots: (i) track community seeds and update them if changes occur, (ii) expand community seeds to complete communities.
\end{enumerate}
To easily track and update community seeds, CUT builds up a Clique Adjacent Bipartite graph (CAB). 
Custom, policies are then introduced to handle node/edge join/removal and seed community expansion.
\\ \ \\
In \cite{Xie2013}, the LabelRankT algorithm is introduced. 
It is an extension of LabelRank \cite{xie2013labelrank}, an online distributed algorithm for the detection of communities in large-scale dynamic networks through stabilized label propagation. 
LabelRankT takes advantage of the partitions obtained in previous snapshots for inferring the dynamics in the current one.
\\ \ \\
The method described in \cite{Lee2014} aims at monitoring the evolution of a graph applying a fading time window. 
The proposed approach maintains a skeletal graph that summarizes the information in the dynamic network. 
Communities are defined as the connected components of such this skeletal graph.
Communities on the original graphs are built by expanding the ones found on the skeletal graph.
\\ \ \\
In \cite{Zakrzewska2015}, the authors propose an algorithm for dynamic greedy seed set expansion, which incrementally updates the traced community as the underlying graph changes. 
The algorithm incrementally updates a local community starting from an initial static network partition (performed trough a classic set seed expansion method). 
For each update, the algorithm modifies the sequence of community members to ensure that corresponding fitness scores are increasing. 
\\ \ \\
In \cite{RPPG15}, the authors introduce TILES\footnote{TILES code available at: \url{https://goo.gl/zFRfCU}}, an online algorithm that dynamically tracks communities in an edge stream graph following local topology perturbations. 
Exploiting local patterns and constrained label propagation, the proposed algorithm reduces the computation overhead. 
Communities are defined as two-level entities:
\begin{itemize}
    \item \emph{Core node}: a core node is defined as a node involved in at least a triangle with other core nodes of the same community;
    \item \emph{Community Periphery}: the set of all nodes at one-hop from the community core.
\end{itemize}
Event detection is also performed at runtime (Birth, Split, Merge, Death, Expansion, Contraction). 

\subsection{Informed CD by network smoothing}
\label{subcat7}

In \cite{Kim2009}, the authors propose a particle-and-density based method, for efficiently discovering a variable number of communities in dynamic networks. 
The method models a dynamic network as a collection of particles called \emph{nano-communities}, and model communities as a densely connected subset of such particles. 
Each particle, to ensure the identification of dense substructures,  contains a small amount of information about the evolution of neighborhoods and/or communities and their combination, both locally and across adjacent network snapshots (which are connected as t-partite graphs). 
To allow flexible and efficient temporal smoothing, a cost embedding technique -- independent on both the similarity measure and the clustering algorithm used -- is proposed. 
In this work is also proposed a mapping method based on information theory. 
\\ \ \\
The method proposed in \cite{Guo2014} finds communities in each dynamic weighted network snapshot using modularity optimization.
For each new snapshot, an input matrix -- which represents a tradeoff between the adjacency matrix of the current step and the adjacency matrix of the previous step -- is computed. 
Then, a custom modularity optimization community detection algorithm is applied.
\\ \ \\
In \cite{Xu2013b,xu2013}, the \emph{Cumulative Stable Contact} (CSC) measure is proposed to analyze the relationship among nodes: upon such measure, the authors build an algorithm that tracks and updates stable communities in mobile social networks. 
The key definition provided in this work regards:
\begin{itemize}
    \item \emph{Cumulative Stable Contact}: There is a CSC between two nodes iff their history contact duration is higher than a threshold.
    \item \emph{Community Core Set}: The community core at time $t$ is a partition of the given network built upon the \emph{useful} links rather than on all interactions.    
\end{itemize}
The network dynamic process is divided into snapshots. 
Nodes and their connections can be added or removed at each snapshot, and historical contacts are considered to detect and update \emph{Community Core Set}. 
Community cores are tracked incrementally allowing to recognize evolving community structures.


\section{Cross-Time Communities Discovery}
Here are collected and classified all the works introducing approaches falling in the \emph{Cross-Time} class.
For details on the class definition and peculiarities refer to Section \ref{subsec:crosstime}.

\subsection{Fixed Memberships, fixed properties}
\label{subcat8}

GraphScope \cite{Sun2007} is an algorithm designed to find communities in dynamic bipartite graphs. 
Based on the Minimum Description Length (MDL) principle, it aims is to optimize graph storage while performing community analysis. 
Starting from an SN, it reorganizes snapshots into segments. 
Given a new incoming snapshot it is combined with the current segment if there is a storage benefit, otherwise, the current segment is closed, and a new one is started with the new snapshot.
The detection of communities sources and destination for each segment is done using  MDL. 
\\ \ \\
In \cite{Duan2009} the authors introduce a method called Stream-Group designed to identify communities on dynamic directed weighted graphs (DDWG). 
Stream-Group relies on a two-step approach to discover the community structure in each time-slice:
\begin{enumerate}
    \item The first step constructs compact communities according to each node's single compactness -- a measure which indicates the degree a node belongs to a community in terms of the graph's relevance matrix.
    \item In the second step, compact communities are merged along the direction of maximum increase of the modularity.
\end{enumerate} 
A measure of the similarity between partitions is then used to determine whether a change-point appears along the time axis and an incremental algorithm is designed to update the partition of a graph segment when adding a new arriving graph into the graph segment. 
In detail, when a new time-slice arrives,
\begin{enumerate}
    \item the community structure of the arriving graph is computed;
    \item the similarity between the partition of the new arriving graph and that of the previous graph segment is evaluated;
    \item change-point detection is applied: if the time-slice $t$ is not a change point, then
    \item the partition of the graph is updated; otherwise a new one is created.
\end{enumerate}

In \cite{Aynaud2011}, the authors propose a definition of average modularity $Q_{avg}$ over a set of snapshots (called time window), defined as a weighted average of modularity for each snapshot, with weights defined \textit{a priori} -- for instance to reflect heterogeneous durations. 
Optimizing average modularity yield a constant partition relevant overall considered snapshots.
This approach does not contemplate community operations.

The authors propose two methods to optimize this modularity:
\begin{itemize}
    \item \emph{Sum-method}: Given an evolving graph $G = \{G_1,G_2, \dots,G_n\}$ and a time window $T \subseteq \{1, \dots,n\}$, a cumulative weighted graph is built, called the sum graph, which is the union of all the snapshots in $T$. Each edge of the sum graph is weighted by the total time during which this edge exists in $T$. 
    Since the sum graph is a static weighted graph, the authors apply the Louvain method \cite{blondel2008fast} on it. This method is not strictly equivalent to optimizing $Q_{avg}$ but allows a fast approximation.
    \item \emph{Average-Method}: two elements of the Louvain method \cite{blondel2008fast} are changed to optimize the average modularity during a time window $T$: (1) the computation of the quality gain in the first phase and (2) how to build the network between communities in the second one.
        \begin{enumerate}
            \item average modularity gain is defined as the average of the static gains for each snapshot of $T$;
            \item given a partition of the node set, the same transformation of Louvain is applied to every snapshot of $T$ independently (with different weights for each snapshot) to obtain a new evolving network between the communities of the partition. 
        \end{enumerate}
\end{itemize}
The authors finally propose a method using sliding windows to segment the evolution of the network in stable periods, potentially hierarchically organized.

\subsection{Fixed memberships, evolving properties}
\label{subcat9}

In \cite{gauvin2014detecting}, the authors propose to use a Non-Negative tensor factorization approach. 
First, all adjacency matrices, each corresponding to a network snapshot, are stacked together in a 3-way tensor (a 3-dimensional matrix with a dimension corresponding to time). 
A custom non-negative factorization method is applied, with the number of desired communities as input, that yields two matrices: one corresponds to the membership weight of nodes to components, and the other to the activity level of components for each time corresponding to a snapshot. 
It can be noted that each node can have a non-zero membership to several (often all) communities. 
A post-process step is introduced to discard memberships below a given threshold and, thus, simplify the results.
\\ \ \\
In \cite{matias2018semiparametric, matias2015estimation} the authors propose to use a Poisson Process Stochastic Block Model (PPSBM) to find clusters in temporal networks.
As in a static SBM, nodes belong to groups, and each pair of groups is characterized by unique connection characteristics. 
Unlike static SBM, such connection characteristic is not defined by a single value, but by a function of time $f$ (intensity of conditional inhomogeneous Poisson process) representing the probability of observing interactions at each point in time. 
The authors propose an adapted Expectation-Maximization (EM) to find both the belonging of nodes and $f$ that better fit the observed network.
It is to be noted that nodes do not change their affiliation using this approach, but instead, it is the clusters' properties that evolve along time.

\subsection{Evolving memberships, fixed properties}
\label{subcat10}

In \cite{Yang2009,Yang2010}, the authors propose to use a Dynamic Stochastic Block Model. 
As in a typical static SBM, nodes belong to clusters, and an interaction probability $\beta_{ql}$ is assigned to each pair of clusters $(q,l)$.
The dynamic aspect is handled by a transition matrix, that determines the probability for nodes belonging to a cluster to move to each other cluster at each step. 
The optimal parameters of this model are searched for using a custom Expectation-Maximization (EM) algorithm. 
The $\beta_{ql}$ are constant for the same community for all time steps.

Two versions of the framework are discussed:
\begin{itemize}
    \item \emph{online learning} which updates the probabilistic model iteratively (in this case, the method is a Temporal Trade-Off CD);
    \item \emph{offline learning} which learns the probabilistic model with network data obtained at all time steps (in this case, the method is a Cross-Time CD).
\end{itemize}
Variations having a similar rationale have been proposed by different authors, most notably in \cite{ishiguro2010dynamic,herlau2013modeling}. The version in \cite{xu2014dynamic} allows properties to evolve, see appropriate section.
\\ \ \\
In \cite{matias2016}, the authors propose to use a Dynamic Stochastic Block Model, similar to \cite{Yang2009}. 
As in a typical static SBM, nodes belong to clusters, and an interaction probability $\beta_{ql}$ is assigned to each pair of clusters $(q,l)$.
The dynamic aspect is handled as an aperiodic stationary Markov chain, defined as a transition matrix $\pi$, characterizing the probability of a node belonging to a given cluster $a$ at time $t$ to belong to each one of the other clusters at time $t+1$. 
The optimal parameters of this model are searched for using a custom Expectation-Maximization (EM) algorithm.
The authors note that it is not possible to allow the variation of $\beta$ and node memberships parameters at each step simultaneously, and therefore impose a single value for each intra-group connection parameter $\beta_{qq}$ along the whole evolution -- thus, searching for clusters with stable intra-groups interaction probabilities. 
Finally, authors introduce a method for automatically finding the best number of clusters using Integrated Classification Likelihood (ICL), and an initialization procedure allowing to converge to better local maximum using an adapted k-means algorithm.
\\ \ \\
In \cite{ghasemian2016detectability}, the authors also use a Dynamic Stochastic Block Model similar to \cite{Yang2009}, and propose scalable algorithms to optimize the parameters of the model based on belief propagation and spectral clustering. They also study the detectability threshold of the community structure as a function of the rate of change and the strength of communities.
\subsection{Evolving memberships, evolving properties}
\label{subcat11}

In \cite{JRF07}, the authors propose to add edges between nodes in different snapshots, i.e. edges linking nodes at $t$ and $t+1$. 
Two types of these edges can be added: 
\begin{itemize}
	\item \textit{identity edges} between the same node, if present at $t$ and $t+1$, and
	\item \textit{transversal edges} between different nodes in different snapshots.
\end{itemize}
To do so, the following relation must hold: there is an edge between $u \in t$ and $v \in t+1$ if $\exists w$ such that $(u,w) \in t$ and $(v,w) \in t+1$. 
A static CD algorithm, Walktrap, is consequently applied to this transversal network to obtain dynamic communities.
\\ \ \\
The method introduced in \cite{Mucha2010} has been designed for any multi-sliced network, including evolving networks. 
The main contribution lies in a multislice generalization of modularity that exploits laplacian dynamics. 
This method is equivalent to adding links between same nodes in different slices, and then run a community detection on the resulting graph. 
In the article, edges are added only between same nodes in adjacent slices. The weight of these edges is a parameter of the method.
The authors underline that the resolution of the modularity can vary from slice to slice. In \cite{Bassett2013}, variations of the same method using different null models are explored.
\\ \ \\
In \cite{VLM15}, the notion of clique is generalized to link streams. 
A $\Delta$-clique is a set of nodes and a time interval such that all nodes in this set are pairwise connected at least once during any sub-interval of duration $\Delta$ of the interval. 
The paper presents an algorithm able to compute all maximal (in terms of nodes or time interval) $\Delta$-cliques in a given link stream, for a given $\Delta$. A solution able to reduce the computational complexity has been proposed in \cite{HMNS16}.
\\ \ \\
In \cite{xu2014dynamic}, a dynamic SBM in introduced. Contrary to \cite{Yang2009}, it allows community properties (i.e., densities between pairs of nodes) to evolve. Note that a potential identifiability problem with allowing both membership and properties to evolve in dynamic SBM was later pointed out in \cite{matias2016}.

\end{document}